\documentclass[aps,twocolumn,showpacs,prc,citeonline,superscriptaddress,amsfonts,amsmath,amssymb,floatfix]{revtex4-1}

\usepackage{graphicx}
\usepackage{amsfonts} 

	\usepackage{amsfonts}
	\usepackage{amssymb}
	\usepackage{subfigure}
	\usepackage{float}

\begin{document} 

\title{Real-space dynamical mean-field theory of Friedel oscillations  in strongly correlated electron systems}

\author{B. Chatterjee}
\affiliation{Institute of Physics, Czech Academy of Sciences, Na Slovance 2, 182 21 Prague, Czech Republic}
\author{J. Skolimowski}
\affiliation{Jo\v{z}ef Stefan Institute, Jamova 39, SI-1000 Ljubljana, Slovenia}
\author{K. Makuch}
\affiliation{Institute of Physical Chemistry, Polish Academy of Sciences, Kasprzaka 44/52, 01-224  Warszawa, Poland}
\author{K. Byczuk}
\email{byczuk@fuw.edu.pl}
\affiliation{Institute of Theoretical Physics, Faculty of Physics, University of Warsaw, ul. Pasteura 5,  02-093 Warszawa, Poland}

\date{\today}

\begin{abstract}
We study Friedel oscillations  and screening effects of the impurity potential  in  the Hubbard model. 
Electronic correlations are accounted for by solving the real-space dynamical mean-field theory equations using the continuous time quantum Monte-Carlo simulations at finite temperatures and using a homogeneous self-energy approximation with  the numerical renormalization group at zero temperature. 
We  find that in the Fermi liquid phase both the amplitudes of Friedel oscillations  and the screening charge decrease with increasing the interaction and follow the behavior  of the Fermi liquid renormalization factor. 
Inside the Mott insulator regime the Friedel oscillations are absent but the residual screening charge remains finite. 
\end{abstract}

\maketitle 

\section{Introduction}
\label{intro}

The Friedel oscillation (FO) is a quantum mechanical phenomenon observed in metals  in  presence of an inhomogeneous, space dependent potential \cite{Friedel52,Friedel58,Kittel,Garcia71}. 
FOs are seen in such physical properties as a charge density or a local density of states at a given energy. 
For example, if either an ion or a defect is inserted into a metal, breaking a translational invariance, then the electrons with energies near the Fermi surface are scattered yielding  spatial oscillations in the charge density surrounding the inhomogeneity.
This effect survives even at finite temperatures $T$  as long as $T/T_{F}\ll1$, where  $T_{F}$ is the Fermi temperature.
FOs can be viewed as a quantum analog of classical screening of a charged impurity in a plasma. 
However, in the classical case the change in the charge density decays exponentially with a distance from the impurity and without oscillations.
 In the quantum regime the decay is polynomial with a rippling pattern formed by  alternating  regions of higher and lower electronic densities. 
The charge density $\bar{n}(r)$ very far from the impurity in a non-interacting, continuous system  is given by 
	\begin{equation}
	 \bar{n}(r)=\bar{n}_{\rm hom}+A \frac {cos(2k_{F}r+\delta)}{r^{d}} ,
	 \label{friedel}
	\end{equation}
 where $\bar{n}_{\rm hom}$ is the uniform density, $ A$ is the amplitude of the FO, $\delta$ is a phase shift, and d is the dimension of the system \cite{Friedel52,Friedel58,Kittel,Garcia71}.
The formula (\ref{friedel}) is valid in the asymptotic limit with $k_{F}r\gg 1$, where $k_{F}$ is the Fermi momentum and $r$ is a  distance from the impurity.
    
 FOs were predicted by  Friedel in a series of papers \cite{Friedel52,Friedel58} and then became an essential component in the study of metallic alloys; for reviews see: \cite{Villain16,Daniel16,Benta16,Georges16}. 
The first direct observations of FOs \cite{Eigler90,Crommie93,Heller94,Manoharan00} were possible after invention of the scanning tunelling microscopy \cite{Binnig82}. 
In fact, signatures of FOs were experimentally seen in various metals and semiconductors, for example, on surfaces Cu$(111)$,  GaAs$(111)$, or Si$(111)$Ag \cite {Eigler90,Kanisawa01,Hasegawa07}. 
In recent years observation of FOs reveals interesting features in solids, such as direction dependent giant charge oscillations on ferromagnetic Fe film grown on W$(001)$ surface \cite{Lounis14} or an interaction induced  band with a well-defined dispersion existing in addition to  conventional surface-state bands at nobel metal surfaces \cite{Bode15}.

After the initial prediction \cite{Friedel52,Friedel58} further theoretical investigations were performed to understand  FOs  in crystals with  non-spherical Fermi surfaces \cite{Clogston62,Callaway64,Adawi66,Rudnick73,Gabovich78,Flores79,Mahan95,Lounis11}, in model systems with an impurity on which the electrons can interact \cite{Sinha70,Affleck08,Tao12,Mitchell15,Derry15}, in one dimensional quantum wires where interacting electrons form a Luttinger liquid \cite{Egger95,Saleur96,Egger97,Noak98,Naon01,Giuliani05,Urban08}, in two or three-dimensonal Hubbard models with electrons forming either the Fermi liquid or the Mott insulator \cite{Ziegler98,Freericks04,Lederer08,Andrade10,Mross11,Chatterjee15}, or in system with interacting and disordered electrons such as amorphous alloys and quasicrystals \cite{Kroha95}. 
Recent theoretical progress was made in understanding spin dependent FOs \cite{Stanescu06}, as well as understanding FOs in topological insulators \cite{Tran10,Liu14}, in graphane \cite{Cheianov06,Hwang08,Virtosztek10}, in cold atoms \cite{Riechers17}, in systems with charge density waves \cite{Vanyolos07} or with presence of the transport currents \cite{Gorczyca07}. 
It was also noted that FOs can be used as a probing tool of quasiparticles \cite{Dalla16}. 
Additionally, FOs lead to an effective interaction between localized magnetic moments which is mediated by conducting electrons and is  known as the  RKKY interaction  \cite{Ruderman54,Kasuya56,Yosida57,Titvinidze12}. 
A related issue to the FOs is the Friedel sum rule which holds for both noninteracting \cite{Friedel52,Friedel58,Rudnick73,Mahan95} as well as for interacting electrons \cite{Langer61,Langreth66,Martin82,Byczuk18}.

Different studies of FOs in  Fermi liquids revealed that the oscillations are renormalized due to the electronic interactions.
Though, the charge FOs in the Hubbard model and the  spin liquid at the Mott transition were studied \cite{Ziegler98,Freericks04,Lederer08,Andrade10,Mross11,Chatterjee15}, a comprehensive quantitative analysis of the FOs at the Mott transition is still an open problem. 
In particular, we would like to address quantitatively  the questions:  How does the oscillation amplitude $A$ change with the interaction strength? How does  it behave at the Mott transition? What is a relation between the Fermi liquid renormalization factor and the amplitude $A$? What is a role of the system dimensionality? 
These problems are particularly interesting in the context of transition metal oxides. It motivated us to study the FOs  in a model system of interacting lattice fermions within the real-space dynamical mean-field theory (R-DMFT) \cite{Dobrosavljevic97,Potthoff99,Freericks04,Okamoto04,Helmes08,Snook08,Byczuk08}. 
The R-DMFT is a reliable, self-consistent and comprehensive approximation for interacting lattice fermions describing both the Fermi liquid as well the Mott insulator \cite{Metzner89,Georges96}.

The paper is organized as follows: In Sec.~\ref{Model} we describe our model.
In Sec.~\ref{formalism} we introduce R-DMFT formalism, discuss methods to solve it in different temperature regimes, and define interesting physical quantities.
In Secs.~\ref{results1}~and~\ref{results2} we present numerical results for FOs as well as  discuss and explain physical  properties of the system.
In Sec.~\ref{summary} we present  conclusions and an outlook for  possible future investigation.

\section{Model}
\label{Model}
 We study  FOs within a one-band Hubbard model in presence of an external impurity potential  and given by
	\begin{equation}
	 H = \sum_{ij \sigma} t_{ij}\  \hat{a}_{i\sigma}^\dag\ \hat{a}_{j\sigma} +\sum_{i\sigma} V_{i\sigma}\ \hat{a}_{i\sigma}^\dag\ \hat{a}_{i\sigma} + U \sum_{i} \hat{n}_{i\downarrow}\hat{n}_{i\uparrow}  , 
	\label{hubbard}
	\end{equation}
    where $\hat{a}_{i\sigma}$ ($\hat{a}_{i\sigma}^{\dag}$) is the annihilation (creation) fermionic operator with spin $\sigma$ on the $i^{th}$ lattice site, $t_{ij}$ is the hopping matrix element between the $i^{th}$ and $j^{th}$ sites with $t_{ii}=0$. 
The second term describes the external (inhomogeneous) potential energy  $V_{i\sigma}$. 
The third term models  the  interaction energy when two fermions with opposite spins are located on the same lattice site.
    
    In this paper we consider a local impurity potential, $V_{i}=V_{0} \ \delta_{i, i_{0}}$, where $i_{0}$ represents the lattice site $\vec{R}_{i_0}$. 
We study one-dimensional (1d)  and two-dimensional (2d) lattices with a  cubic type  geometry. 
We consider  paramagnetic systems without any  long-range order.
The Hamiltonian (\ref{hubbard}) is solved within  R-DMFT approximation \cite{Potthoff99,Freericks04,Okamoto04,Helmes08,Snook08,Byczuk08}.

\section{R-DMFT formalism and physical quantities}
\label{formalism}
\subsection{Matsubara Green's functions}

 All physical properties, which are studied here, are obtained from  one-particle Green's functions defined by 
    \begin{equation}
      G_{ij\sigma}(\tau) = - \langle T_{\tau} \hat{a}_{i\sigma}(\tau) \hat{a}_{j\sigma}^\dag (0)\rangle,
      \label{greenfunction}
    \end{equation}
 where  $\tau \in (0,\beta)$ is the imaginary time and $\beta=1/T$ denotes the inverse temperature with the Boltzmann constant equal to unity \cite{Negele88}. 
The symbol $T_{\tau}$ means the chronological operator and 
$\langle...\rangle$ represents both quantum and  thermal averages in the grand-canonical ensemble with a fixed chemical potential $\mu$.
 Later we also perform  Fourier transformation to obtain the Green's functions $ G_{ij\sigma}(i\omega_{n})$, where $\omega_{n}=(2n+1)\pi/\beta$ are fermionic Matsubara frequencies with integer $n$.

    \subsection{R-DMFT}

 The Green's functions (\ref{greenfunction}) in Matsubara frequency space obey an exact Dyson equations
    \begin{equation}
     G_{ij\sigma} (i\omega_{n})=  G_{ij\sigma}^{(0)} (i\omega_{n})+ \sum_{kl} G_{ik\sigma}^{(0)} (i\omega_{n})\Sigma_{kl\sigma}G_{lj\sigma} (i\omega_{n}),
     \label{Dyson}
    \end{equation}
  where $G_{ij\sigma}^{(0)} (i\omega_{n})$ are the Green's functions at $U=0$.
 The self-energies $\Sigma_{kl\sigma} (i\omega_{n})$ account for all interaction effects.
    The main approximation of R-DMFT \cite{Metzner89}  is that the self-energies are local, which means that they are diagonal in the lattice site indices, i.e. 
    \begin{equation}
     \Sigma_{ij\sigma} (i\omega_{n})= \Sigma_{i\sigma}(i\omega_{n})\delta_{ij},
      \label{dmftse}
     \end{equation}
   where $\delta_{ij}$ is the Kronecker delta. 
Nevertheless, they are site dependent for inhomogeneous systems \cite{Potthoff99,Freericks04,Okamoto04,Helmes08,Snook08,Byczuk08}. 

 R-DMFT approximation consists of the following  set of self-consistent equations: 
   At each lattice site $l$ the reduced partition function, obtained within  the cavity method \cite{Georges96}, is given by
   \begin{equation}
    Z^{l}=\int D [a_{l\sigma},a_{l\sigma}^{\ast}] e^{-S_{l}[a_{l\sigma},a_{l\sigma}^{\ast}]},
   \end{equation}
    where  the local action $S^l$  is  
     \begin{align}
    S_{l} & =-\int\limits_{0}^{\beta}d\tau_{1}\int\limits_{0}^{\beta}d\tau_{2}\sum_{\sigma}a_{l\sigma}^{\ast}(\tau_{1})\mathcal{G}_{l\sigma}^{-1}(\tau_{1}-\tau_{2})a_{l\sigma}(\tau_{2})\nonumber\\
   & +U \int\limits_{0}^{\beta}d\tau a_{l\uparrow}^{\ast}(\tau)a_{l\uparrow}(\tau)a_{l\downarrow}^{\ast}(\tau) a_{l\downarrow}(\tau).
   \label{locacc3}
   \end{align}
   In Eq.~(\ref{locacc3}) the path integral formalism in the coherent state representation is used with  $a_{l\sigma}^{\ast}$ and  $a_{l\sigma}$ as  Grassmann variables \cite{Negele88}.
In Eq. (\ref{locacc3}) the kernel 
	 \begin{equation}
    \mathcal{G}_{l\sigma}^{-1}(\tau_{1}-\tau_{2})\equiv -\left( \frac{\partial}{\partial\tau_{1} }-\mu+V_{l}\right)\delta(\tau_{1}-\tau_{2})-\Delta_{l\sigma}(\tau_{1}-\tau_{2})
    \label{weisstau}
   \end{equation}
is the inverse of the mean-field propagator and
    \begin{equation}
    \Delta_{l\sigma}(\tau_{1}-\tau_{2})\equiv -\sum_{ij}t_{li} G_{ij\sigma}^{l} (\tau_{1}- \tau_{2})t_{jl}
    \label{hybtau}
   \end{equation}
   is the hybridization function, which physically accounts for the coupling of the $l^{th}$ site with the rest of the lattice.
 $G_{ij\sigma}^{l}(\tau)$ is the Green's function on a lattice with a cavity on site $l$.
The symbol $\delta(\tau)$ represents the Dirac function. 
 The mean-field propagators $\mathcal{G}_{l\sigma} (i\omega_{n})$ are related to the diagonal matrix elements 
$[{\bf G}_{\sigma}(i\omega_{n})]_{ll}$ of the one-particle matrix Green's function by the local Dyson equation 
   \begin{equation}
    \mathcal{G}_{l\sigma}^{-1}(i\omega_{n})= [{\bf \Sigma}_{ \sigma} (i\omega_{n})]_{ll} + 
\frac{1}{[{\bf G}_{\sigma}(i\omega_{n})]_{ll}}.
    \label{localdyson}
   \end{equation}
   We use notation where the matrices are given by  $[{\bf G}_\sigma(i\omega_{n})]_{ij}=G_{ij\sigma}(i\omega_{n})$ and $[{\bf \Sigma}_{\sigma}(i\omega_{n})]_{ij}= \Sigma_{ij\sigma}(i\omega_{n})$, respectively.   
    The lattice Green's functions are obtained by inverting the real space Dyson equations (\ref{Dyson}), which in a matrix form read  
   \begin{equation}
    {\bf G_{\sigma}}(i\omega_{n})=[{\bf \xi} (i\omega_{n})-{\bf t}-{\bf \Sigma_{\sigma}}(i\omega_{n})]^{-1},
    \label{redyson}
   \end{equation}
   where $[{\bf \xi} (i\omega_{n})]_{ij} = (i\omega_{n}+ \mu -V_{i})\delta_{ij}$.
 The hopping matrix is given by $[{\bf t}]_{ij}=t_{ij}$ and the matrix self-energy  is obtained within the local approximation (\ref{dmftse}).
 Finally, the functional integrals determining the diagonal matrix elements of the Green's functions are given by
    \begin{align}
    [{\bf G}_\sigma (i\omega_{n})]_{ll} & =-\frac{1}{Z^{l}}\int\prod_{\sigma} D [a_{l\sigma}^{\ast}, a_{l \sigma}][a_{l\sigma}(i\omega_{n}) a_{l\sigma}^{\ast} (i\omega_{n})]\nonumber\\
   &  e^{-S_{l}  [a_{l\sigma}^{\ast}, a_{l \sigma}]}.
   \label{partgreen}
   \end{align}
   The set of Eqs.~(\ref{Dyson})-(\ref{partgreen}) constitutes R-DMFT and  those equations are solved numerically in an  iterative way.

 \subsection{R-DMFT within CT-QMC}

Among different R-DMFT self-consistency equations  (\ref{Dyson})-(\ref{partgreen})  the most difficult  is to solve the problem in Eq.~(\ref{partgreen}).
Here we solve it by using continuous time quantum Monte Carlo (CT-QMC) simulations, where the partition function is expanded about the hybridization function and resummed  by using a stochastic Metropolis algorithm \cite{ctqmc}. 
The computer program developed by us  is based on the approach of Haule \cite{Haule07}. 
In this method the problem is solved in the Matsubara frequency space. 
CT-QMC only works for finite temperatures and the computational time scales at least linearly with $\beta$.
 In this method the problem (\ref{partgreen}) is solved on every non-equivalent lattice site. 
These factors made solution of  R-DMFT  within CT-QMC a very time consuming problem, in particular, at low temperatures and for large lattice systems.

  \subsection{R-DMFT within homogeneous self-energy approximation}

Due to these  limitations of  CT-QMC we also use an approximate method to solve R-DMFT equations, which is described now. 
The self-energy $\Sigma_{i\sigma}(i\omega_{n})$ in Eq.~(\ref{Dyson}) can be split into a homogeneous and an inhomogeneous part as follows
	\begin{equation}
	\Sigma_{i\sigma}(i\omega_{n})=\Sigma_{\sigma}(i\omega_{n})+\Delta \Sigma_{i\sigma}(i\omega_{n}).
	\label{selfenergy}
	\end{equation}
The first term  accounts for the interaction effects in a homogeneous system and it is therefore site independent. 
We want to describe  FOs very far from the impurity, i.e.   at sites $\vec{R}_i$ such that $\lvert \vec{R}_{i}-\vec{R}_{i_{0}}\lvert \gg a$, where $a$ is a lattice constant, and make a contact with the original formula (\ref{friedel}). 
Therefore, since far away from the impurity the effect of $V_{i\sigma}$ on the self-energy is expected to be weak  the inhomogeneous part  $\Delta \Sigma_{i\sigma}(i\omega_{n})$ can be neglected. 
The homogeneous part $\Sigma_{\sigma}(\omega)= \Sigma_{\sigma}(i\omega_{n}\longrightarrow\omega+i0^{+})$ is determined by solving the dynamical mean-field theory (DMFT) self-consistency equations  for infinite homogeneous system \cite{Georges96} at zero temperature by using the numerical renormalization group (NRG) method \cite{Bulla08}. 
The open-source code {\em NRG Ljubljana}  is used for that purpose \cite{nrgljubljana}. 
The computed self-energy is then transferred into the real space Dyson equation (\ref{redyson}) containing the impurity potential $V_{i}$ in order to obtain the Green's function.
This approach to solving R-DMFT is called the homogeneous self-energy approximation (HSEA).

We note that HSEA is not equivalent to the commonly used local density approximation (LDA), e.g. \cite{Snook08}, and only the former gives rise to FOs. 
The LDA, also known as a Thomas-Fermi or WKB approximation, corresponds to the replacement of  k-integrated Dyson equation in the homogeneous DMFT \cite{Georges96} by $G_{ii \sigma}(\omega) = \sum_{\vec{k}} 1 /[\omega - \epsilon_{\vec{k}} -V_i -\Sigma(\omega) ]$, where $\epsilon_{\vec{k}}$ is a non-interacting dispersion relation in momentum 
$\vec{k}$-space. Within  HSEA the real-space Dyson equation (\ref{Dyson}), after Fourier transformation, takes the form $G_{ii\sigma}(\omega) = \sum_{\vec{k}\vec{k}'}  e^{i \vec{R}_i (\vec{k}-\vec{k}' )} G_{\vec{k} \vec{k}'}(\omega)$, where $G_{\vec{k} \vec{k}'}(\omega)=[{\bf G}^0(\omega)^{-1}-\Sigma(\omega) {\bf 1}]^{-1}_{\vec{k} \vec{k}' }$, with $[{\bf 1}]_{\vec{k} \vec{k}' }=\delta_{\vec{k} \vec{k}' }$, $[{\bf G}^0(\omega)^{-1}]_{\vec{k} \vec{k}' }=(\omega-\epsilon_{\vec{k}})\delta_{\vec{k} \vec{k}' }-V_{\vec{k} \vec{k}' }$ , and 
$V_{\vec{k}\vec{k}'} $ being  a Fourier transform of the potential $V_i$.  The presence of  oscillatory  terms with $\vec{k}\neq \vec{k}'$ yields  FOs. 
    
\subsection{Physical quantities}
 
The most desired physical quantity in studying the FOs is the average number of particles on each lattice site $ \bar{n}_{i\sigma}\equiv  \langle \hat{a}_{i\sigma}^\dag \hat{a}_{i\sigma}\rangle$, i.e. the spin-resolved particle density. 
In case of  R-DMFT solved within CT-QMC the spin resolved density of particles, given by
\begin{equation} 
\label{local_occupation}
 \bar{n}_{i\sigma} =  \lim_{\tau \to 0^{-}} G_{ii\sigma}(\tau)
\end{equation}
is directly determined from  the Monte Carlo simulations. 
Since this method is based on stochastic sampling, the statistical error given by the standard deviation of  different samples is also estimated here. 
 
In case of HSEA we determine the retarded one-particle Green's function \cite{Negele88}  and then we find the local spectral function 
	\begin{equation}
	A_{i \sigma}(\omega)=-\frac{1}{\pi} \text{Im} \ G_{ii\sigma}(\omega).
	\label{spectral}
	\end{equation}
 Having $A_{i \sigma}(\omega)$  we compute the spin-resolved particle density at finite temperatures according to  
 \begin{equation} \label{local_occupationhom}
   \bar{n}_{i\sigma}= \int_{-\infty}^{+\infty} A_{i\sigma}(\omega)f(\omega) \,d\omega,
   \end{equation}
   where $f(\omega)$ is the Fermi-Dirac distribution function.

The total number of particles per site is given by
	\begin{equation}
	\bar{n}=\frac{1}{N_{L}}\sum_{i=1}^{N_L} \bar{n}_{i},
	\end{equation}
where $\bar{n}_{i}=\bar{n}_{i\uparrow}+\bar{n}_{i\downarrow}$ and $N_{L}$ is the number of the lattice sites.
Since in this paper we consider spin-rotationally invariant systems, the equality  $\bar{n}_{i\uparrow}=\bar{n}_{i\downarrow}$ holds. 

The screening effect, i.e. a shielding of the impurity potential by the conducting electrons \cite{Kittel},  is quantified by the so-called  screening charge according to 
	  \begin{equation}
	  Z=\sum\limits_{i}(\bar{n}_{i}-\bar{n}_{\rm hom}),
	  \label{screen1}
	\end{equation}
    where the summation runs over all  lattice sites and $\bar{n}_{\rm hom}$ corresponds to the particle  density of the corresponding homogeneous system with $V_i=0$.

\section{Numerical results for R-DMFT within CT-QMC}
\label{results1}

    In this Section we present numerical results obtained within CT-QMC solver for R-DMFT equations. 
In the following   we choose the chemical potential $\mu= U/2$ so that the homogeneous system is at half-filling with the density $\bar{n}=1$. 
We consider a linear chain of atoms  and a square lattice. 
In all cases the hopping amplitude $t_{ij}=t$ is only between nearest neighbours. 
We set $t=1$ to define the energy unit and the lattice constant $a=1$ to define the length unit. 
The band-width $W$ is given by $W=2zt$, where $z$ is the co-ordination number. 
The system is subjected to periodic boundary conditions with a finite number $N_{L}$ of the lattice sites. 
In case of R-DMFT within CT-QMC we perform simulations for finite temperatures corresponding to $\beta$ between $5t$ and $20t$.  These temperatures are within the  cross-over regime  of the $U-T$ phase diagram, above the critical temperature for the square lattice \cite{Schafer15}. 

 \subsection{One-dimensional chain}
 
\begin{figure} [ht!]
		    \centering
		      \includegraphics[width=0.5\textwidth]{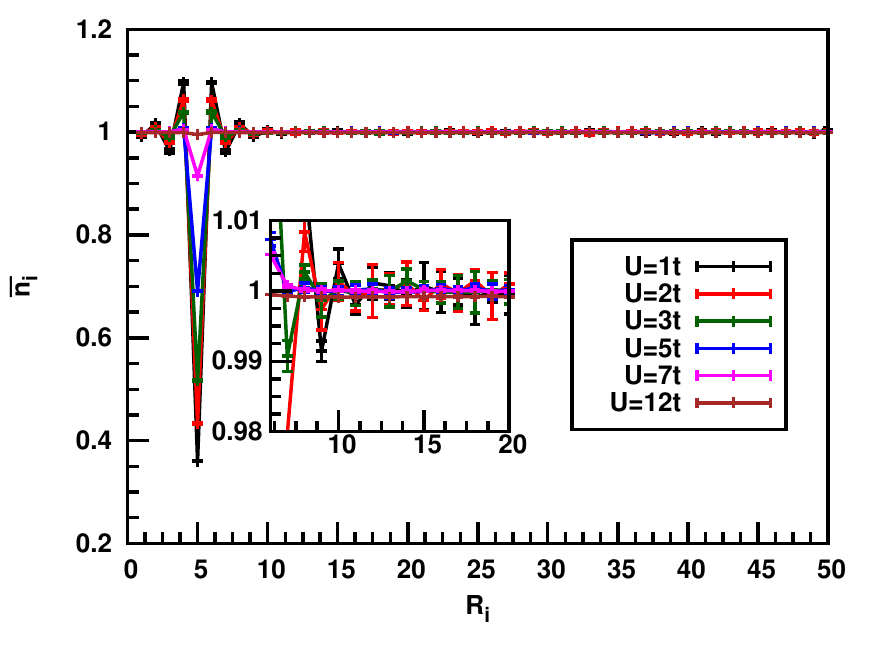} 
		      \qquad
		      \includegraphics[width=0.5\textwidth]{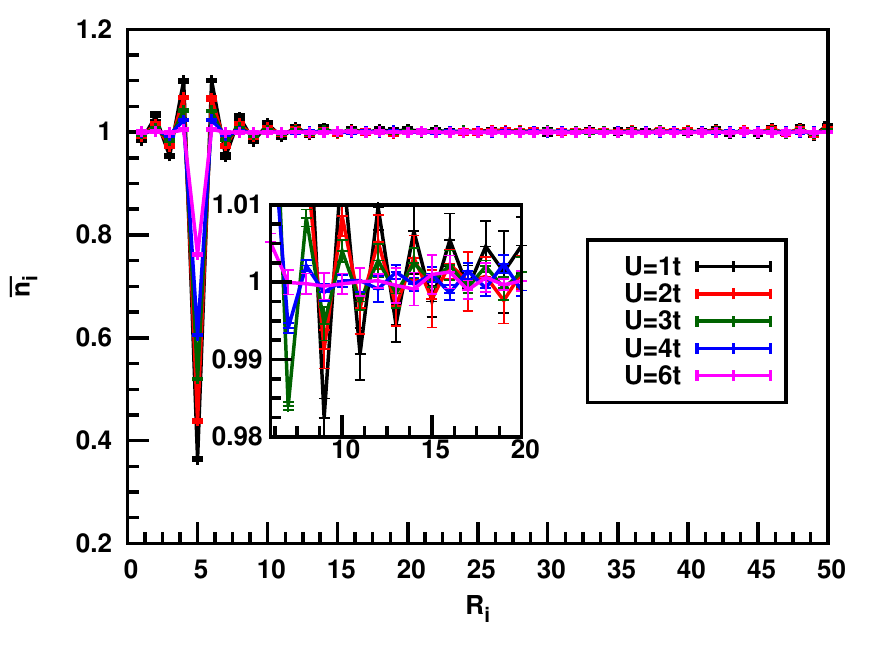}
		      \caption{Friedel oscillations in  particle densities  at different lattice site in presence of the single impurity potential $V_{0}=2t$ placed at $|\vec{R}_i|=5a$ in a 1d chain with $N_{L}=50$ sites. Upper (lower) panel presents results for $\beta=5t$ ($\beta=20t$).  Different colors correspond to different interactions $U$ which is accounted for by using R-DMFT within CT-QMC. Corresponding stochastic error bars are also shown. The insets show FOs in the neighborhood of the impurity.}
		     \label{fig1}
	             \end{figure}

 In Fig.~\ref{fig1}   we present the density $\bar{n}_i$   on different lattice sites where $N_L=50$ and when the impurity potential $V_{0}=2t$ is set on  $|\vec{R}_i|=5a$ site. 
 The upper panel shows FO at $\beta =5t$ and the lower panel shows FO at $\beta=20t$.  
 Results are presented for different interactions $U$ marked by lines in different colors. 
 In the zoomed areas we show FO in the vicinity of the impurity site. 
 Away from the impurity site the relative changes in the densities are very small, much lower than 1\%. 
 As one expects the oscillations are more pronounced at lower temperatures although it was surprising to see them clearly at $\beta=5t$ as well. 
 With increasing $U$ the FOs are weaker and disappear completely when $U > U_c$, where $U_c$ is the critical interaction where the Mott-Hubbard MIT occurs \cite{1D,Lieb68}. 
 We note that  the period and the phase-shift of the oscillations remain unchanged and stay the same at different temperatures. 
 With the help of Eq.~(\ref{friedel}) it means that at half-filling the length of the Fermi wave vector $k_F$ and the phase shift are invariant with respect to the interaction changes \cite{note-phase-shift}.

\begin{figure} [ht!]
		    \centering
		      \includegraphics[width=0.5\textwidth]{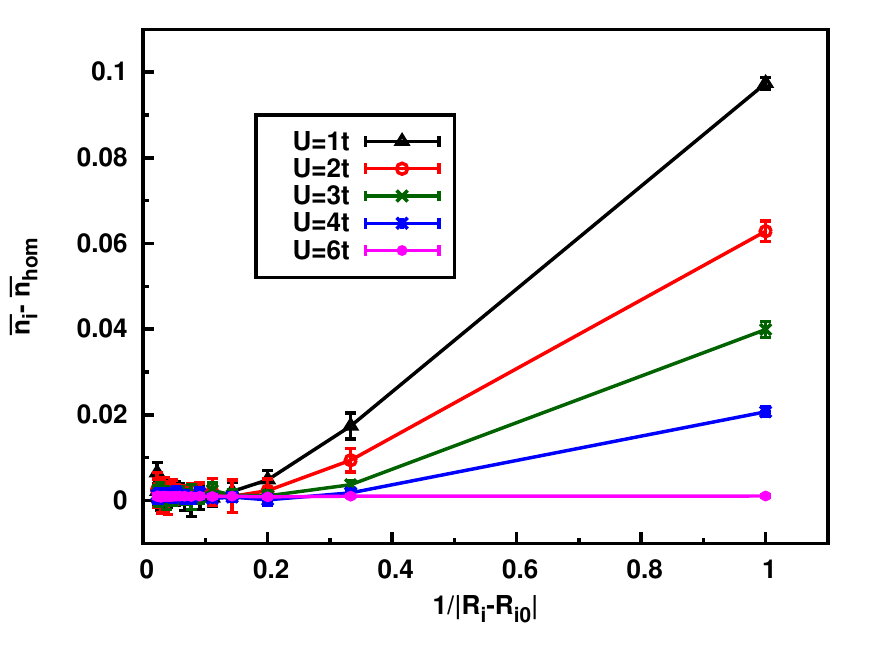} 
		      \qquad
		      \includegraphics[width=0.5\textwidth]{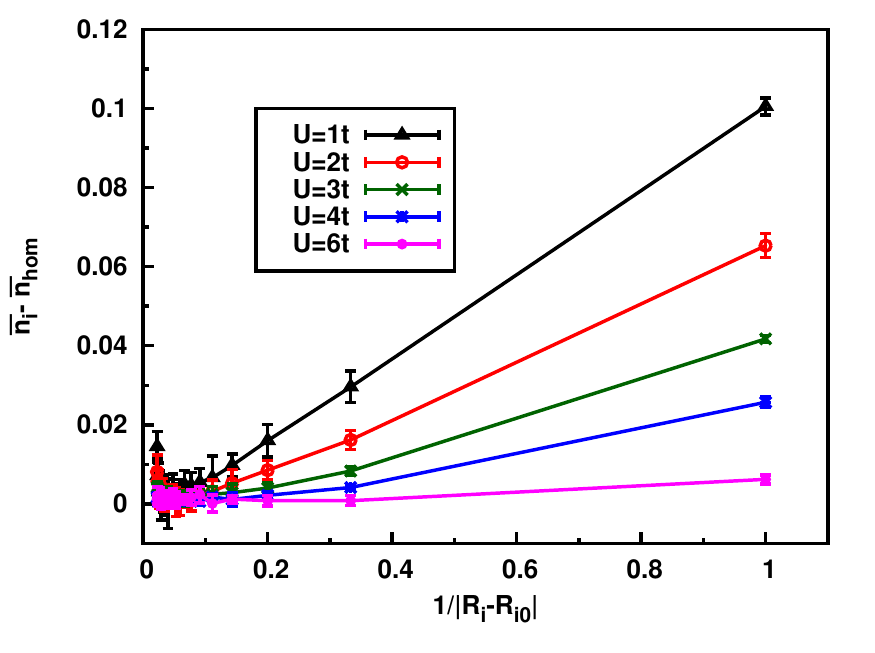}
		      \caption{Variation of the  density deviations as a function of inverse of the relative distance from the impurity site for the same system and method as in Fig.~\ref{fig1}.  In the upper (lower) panel we show results for $\beta=5t$  ($\beta=20t$).}
		     \label{fig2}
		      \end{figure}

\begin{figure} [ht!]
		   		    \centering
			 \includegraphics[width=0.5\textwidth]{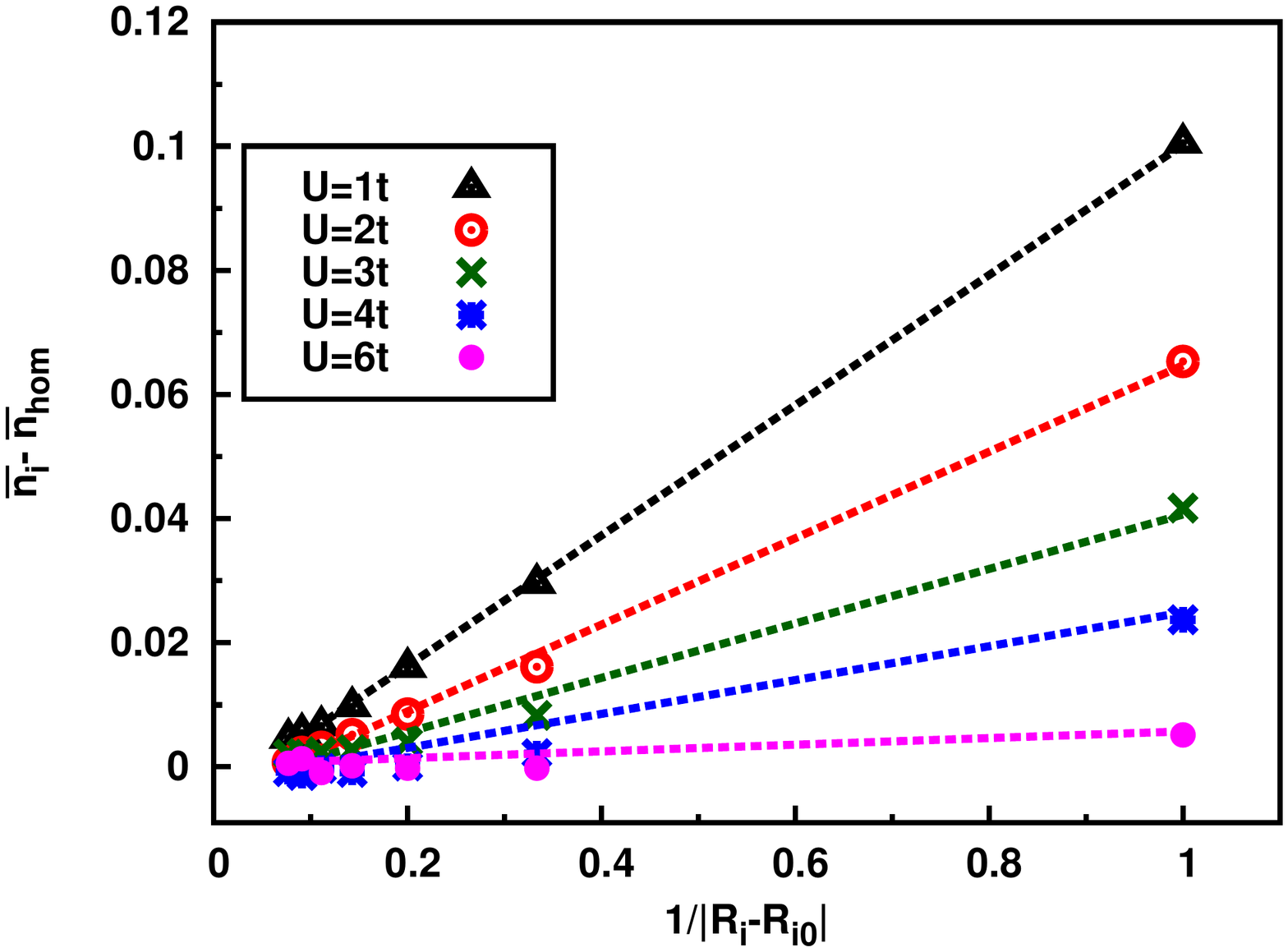} 
			  \caption{Linear fits for points in Fig.~\ref{fig2} in the lower panel for $\beta=20 t$. The slope $A(U)$  decreases with increasing $U$ and vanishes on the insulating side.}
			 \label{fig3}
			  \end{figure}

\begin{figure} [ht!]
		       		    \centering 
			 \includegraphics[width=0.5\textwidth]{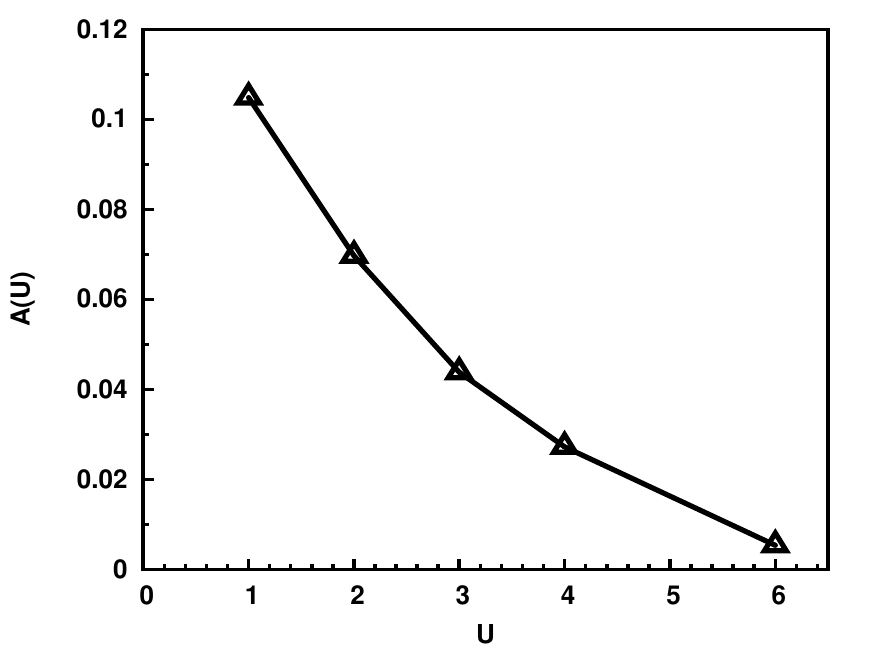} 
			  \caption{Variation  of the slope $A(U)$  determined  from Fig.~\ref{fig3} for $\beta=20t$ as a function of $U$. }
			  \label{fig4}
			  \end{figure}

		      In order to have a more quantitative understanding on the decay of the FO amplitudes with the increasing interaction we plot in Fig.~\ref{fig2} the local density deviations $|\bar{n}_{i}-\bar{n}_{\rm hom}|$ as a function of the inverse of the relative distance from the impurity site, i.e. $1/|\vec{R}_{i}-\vec{R}_{i_{0}}|$. 
		      The asymptotic linear decay is visible in agreement with Eq.~(\ref{friedel}) in the presence of interactions for both temperatures. 
		      In case of $\beta=5t$ we see more random deviations in the maxima of the local density as compared to $\beta=20t$. 
		      This is due to a smallness of the  FO amplitudes in the former case and stochastic nature of the CT-QMC results. 
		      The points shown in Fig.~\ref{fig2}  follow the approximate linear rule $|\bar{n}_{i}-\bar{n}_{hom}|=Ax+B$, where $x=1/|\vec{R}_{i}-\vec{R}_{i_{0}}|$. 
		      The parameters  $A$ and $B$ can be  determined by fitting procedure. 
		      Of particular interest is the slope parameter $A=A(U)$ which describes changes in the FO amplitude with $U$.
		      We perform a linear fit for the case of $\beta=20$ and present it in Fig.~\ref{fig3}. 
		      We see that the slope $A(U)$  decreases with increasing $U$ and vanishes at the metal-insulator transition, cf. Fig.~\ref{fig4}.  
		      It is seen that the  Mott transition occurs at   $U\approx 6t$ for $\beta=20 t$. 
		      We note that the linearized DMFT \cite{Bulla00} predicts $U_c=6\sqrt{z} t\approx 8.5 t$ at $T=0$ so our value at finite $T$ remains in a good agreement.

\begin{figure} [ht!]
\centering
\includegraphics[width=0.5\textwidth]{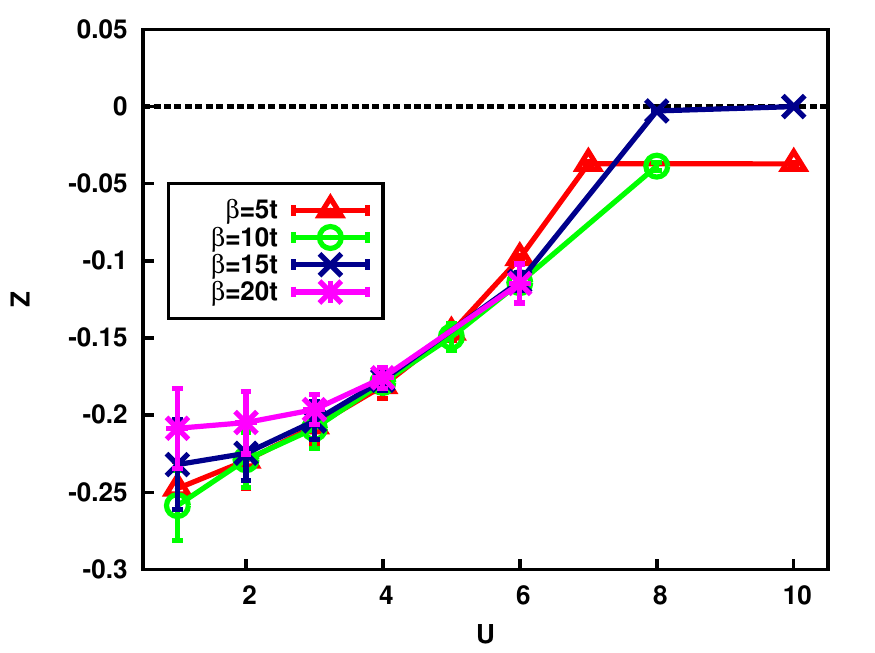} 
\qquad
\includegraphics[width=0.5\textwidth]{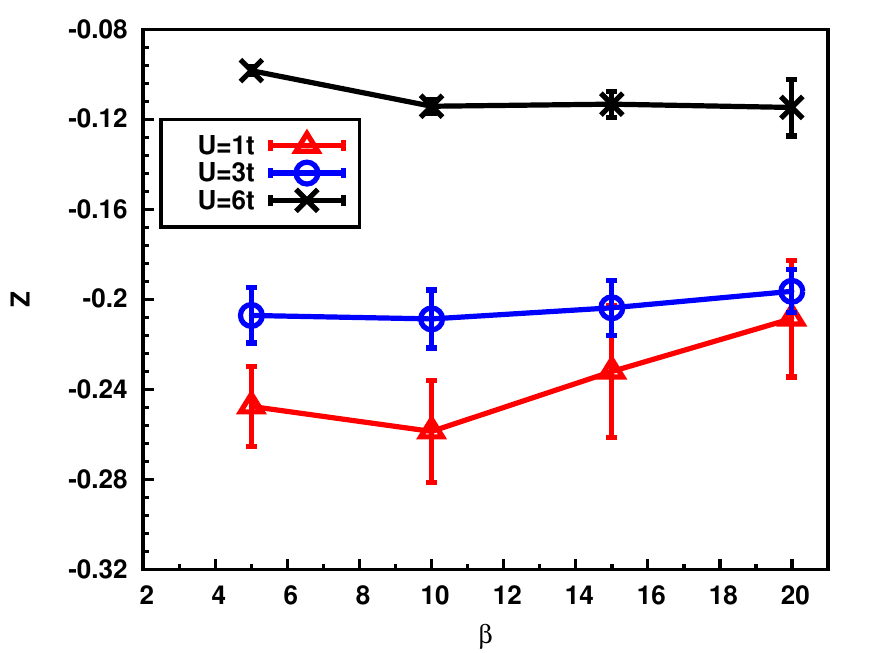} 			 
\caption{Screening charge $Z$ determined from Eq.~(\ref{screen1}) as a function of interaction $U$ (upper panel) and temperature (lower panel). The system parameters are the same as in Fig.~\ref{fig1}. }
\label{fig5}
\end{figure}

The screening charge $Z$, defined in Eq.~(\ref{screen1}), is shown in Fig.~\ref{fig5}. 
Since the impurity potential is repulsive $V_0=2t>0$ the particles are pushed away ($Z<0$) from the system which is treated within the grand canonical ensemble with a constant chemical potential $\mu=U/2$ as discussed at the beginning. 
With increasing the interaction the screening of the impurity is weaker, i.e. the number of removed charge is smaller, as is seen in the upper panel of Fig.~\ref{fig5}. 
When the system turns into a Mott phase with an open correlation gap, the screening is ineffective and $Z$ approaches zero as shown in the  upper panel of Fig.~\ref{fig5}. 
We find rather  weak dependence of the screening charge on the temperature, which is illustrated in the lower panel of Fig.~\ref{fig5}.

\subsection{Two-dimensional square lattice}

\begin{figure} [ht!]
\centering
\includegraphics[width=0.5\textwidth]{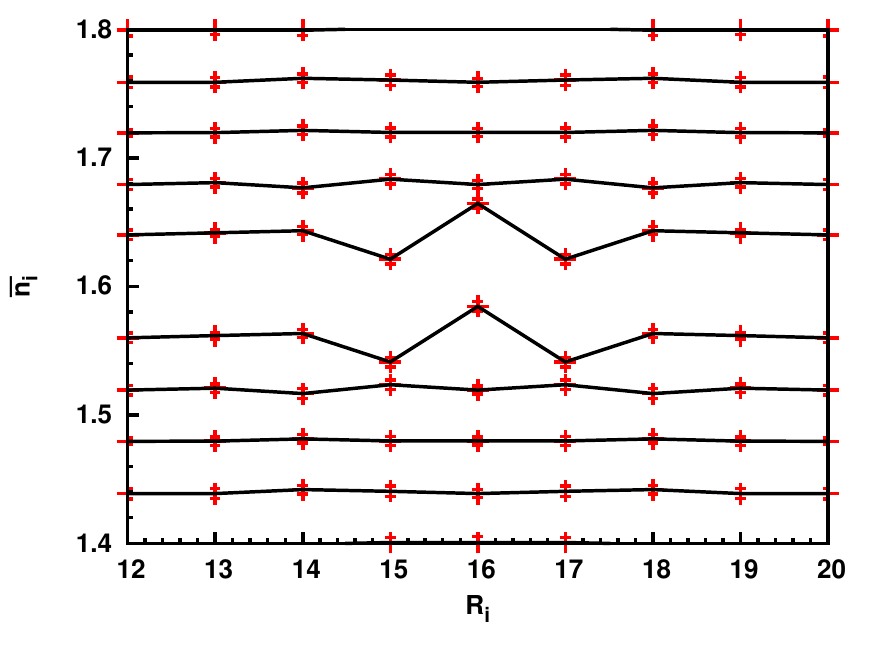} 
\qquad
\includegraphics[width=0.5\textwidth]{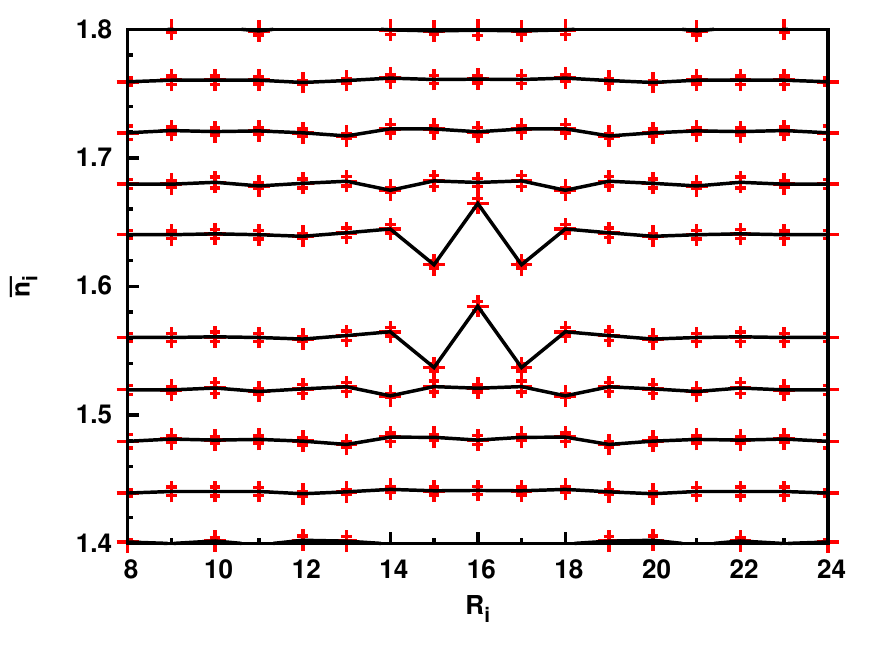} 			 
\caption{ Friedel oscillations in particle densities of the Hubbard model in two-dimensional square lattice with $31\times 31$ sites. The impurity potential $V_0=6t$ is located in the center at ${\bf R}_0=(15a,15a)$ and the interaction $U=t$. Densities are plotted along the horizontal ($x$) axis for different sections in the vertical ($y$) direction with a lateral shift $0.04$. The line with the impurity potential is excluded. In the upper (lower) panel the temperature is set to $\beta=5t$ ($15t$).  Stochastic error bars are shown. }
\label{fig6}
\end{figure}

\begin{figure} [ht!]
\centering
\includegraphics[width=0.5\textwidth]{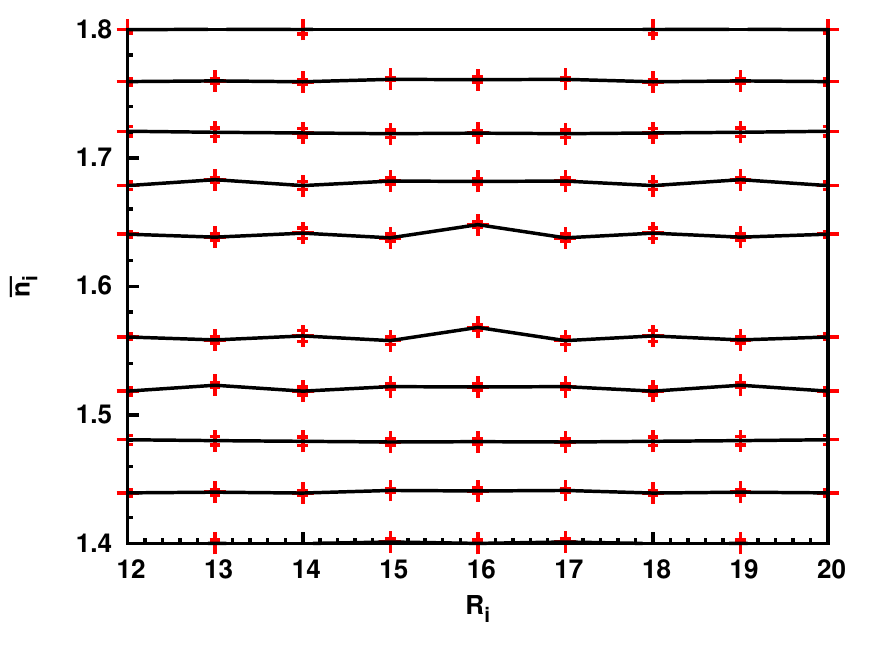} 
\qquad
\includegraphics[width=0.5\textwidth]{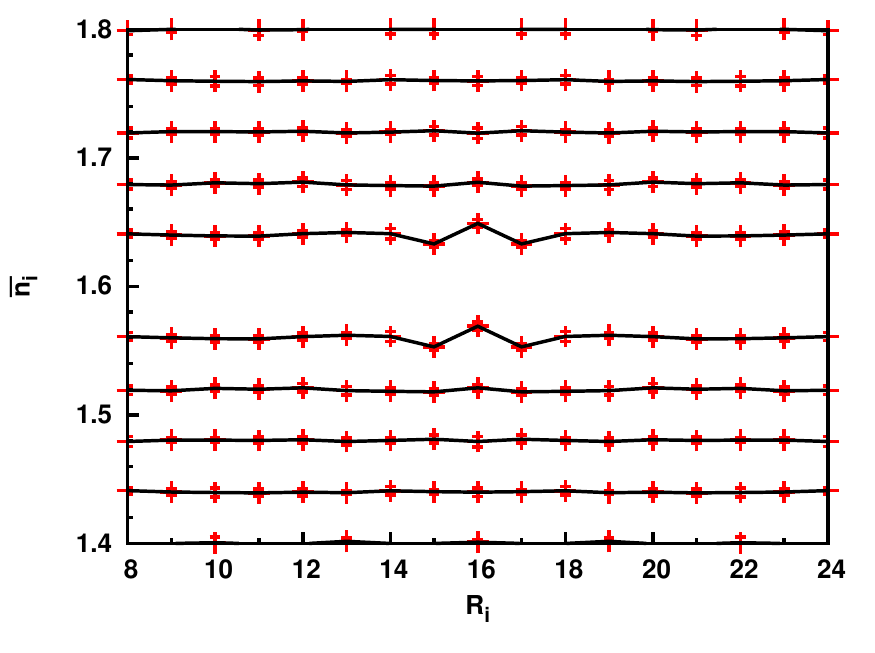} 			 
\caption{ Friedel oscillations in particle densities of the Hubbard model with $U=5t$ in two-dimensional square lattice. The other parameters and style of plotting are the same as in Fig.~\ref{fig6}.}
\label{fig7}
\end{figure}

The FOs in two-dimensional lattice are presented in Figs.~\ref{fig6} and \ref{fig7}. The system size is $31\times 31$  and the impurity  site is in the center at $\vec{R}_0=(15a,15a)$. In Fig.~\ref{fig6} we present results at $U=t$ and in Fig.~\ref{fig7} at $U=5t$. It is clearly seen that with increasing the interaction the FO amplitudes diminish similarly as in one-dimensional case. 

We note that FOs are stronger, i.e. more visible, at lower temperatures, cf. upper and lower panels of Figs.~\ref{fig6} and \ref{fig7}, respectively. 
Since the FO decays as $1/r^2$ in two dimensions, values of the amplitude are much smaller than those in the previous $d=1$ case. 
This fact together with stochastic nature of the CT-QMC makes it impossible to provide more detailed analysis. 
Also getting results for FOs in three dimensional systems is beyond the present CT-QMC method. Therefore we turn to HSEA approach for the further studies of FOs in the Hubbard model.

\section{Numerical results for R-DMFT within HSEA}
\label{results2}

In this Section we present numerical results obtained within HSEA. 
We determine the DMFT self-energy for a given $U$ by using the NRG at zero temperature. 
This self-energy is inserted into the Dyson Eq.~(\ref{Dyson}) to obtain the Green's functions and  other physical quantities, as discussed in Sec.~III. 
Since we start from using the NRG self-energies, the Green's functions are determined on the real-frequency axis. 
As in the former Section, the chemical potential is fixed at $\mu=U/2$ to keep the homogeneous systems at half-filling with $\bar{n}=1$. 
We consider a linear chain of atoms and a square lattice here. 
In both cases the hopping amplitude $t_{ij}=t$ is only between nearest neighbors and $t=1$ as earlier. 

 \subsection{One dimensional chain}
 
\begin{figure} [ht!]
\centering
\includegraphics[width=0.5\textwidth]{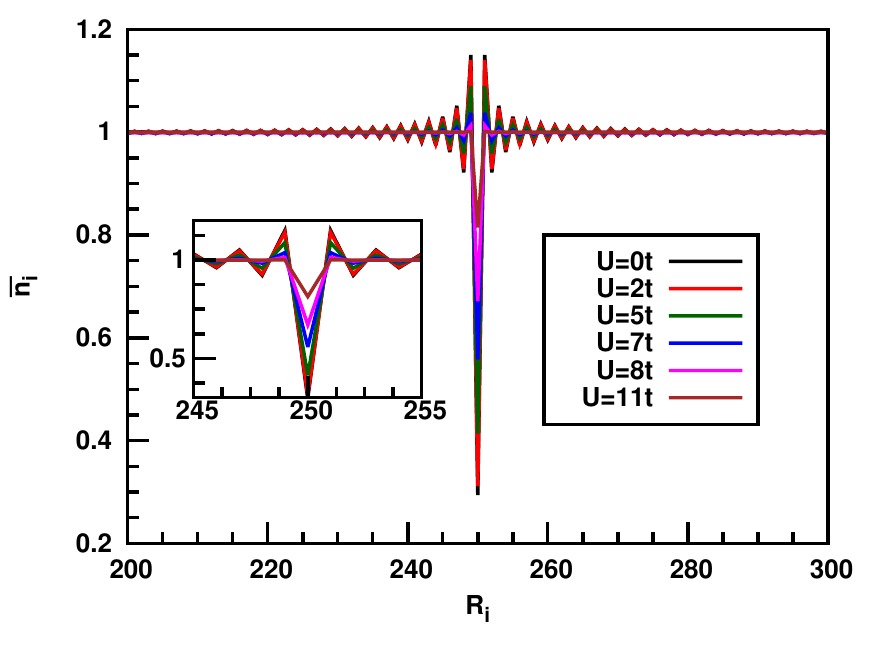} 		 
\caption{ Friedel oscillations in particle densities  at different lattice site in presence of the single impurity potential $V_{0}=2t$ placed at $|\vec{R}_i|=250a$ in a 1d chain with $N_{L}=500$ sites. Different colors correspond to different interactions $U$ which is accounted for by using R-DMFT within HSEA. The inset shows FOs in the neighborhood of the impurity.}
\label{fig8}
\end{figure}

 In Fig.~\ref{fig8}   we present the density $\bar{n}_i$   on different lattice sites where $N_L=500$ and when the impurity potential $V_{0}=2t$ is set on the $|\vec{R}_i|=250a$ site. 
 Within the HSEA we are able to work within much larger lattices as compared to the previous case of CT-QMC. 
 Results are presented for different interactions $U$ marked by lines in different colors. 
 In the zoomed areas we show FO in the vicinity of the impurity site. 
 Away form the impurity site the relative changes in the densities are very small, much lower than 1\% even at $T=0$.  
 With increasing $U$ the FOs are weaker and disappear completely when $U > U_c$, where $U_c$ is the critical interaction where the Mott-Hubbard MIT occurs.
 
\begin{figure} [ht!]
\centering
\includegraphics[width=0.5\textwidth]{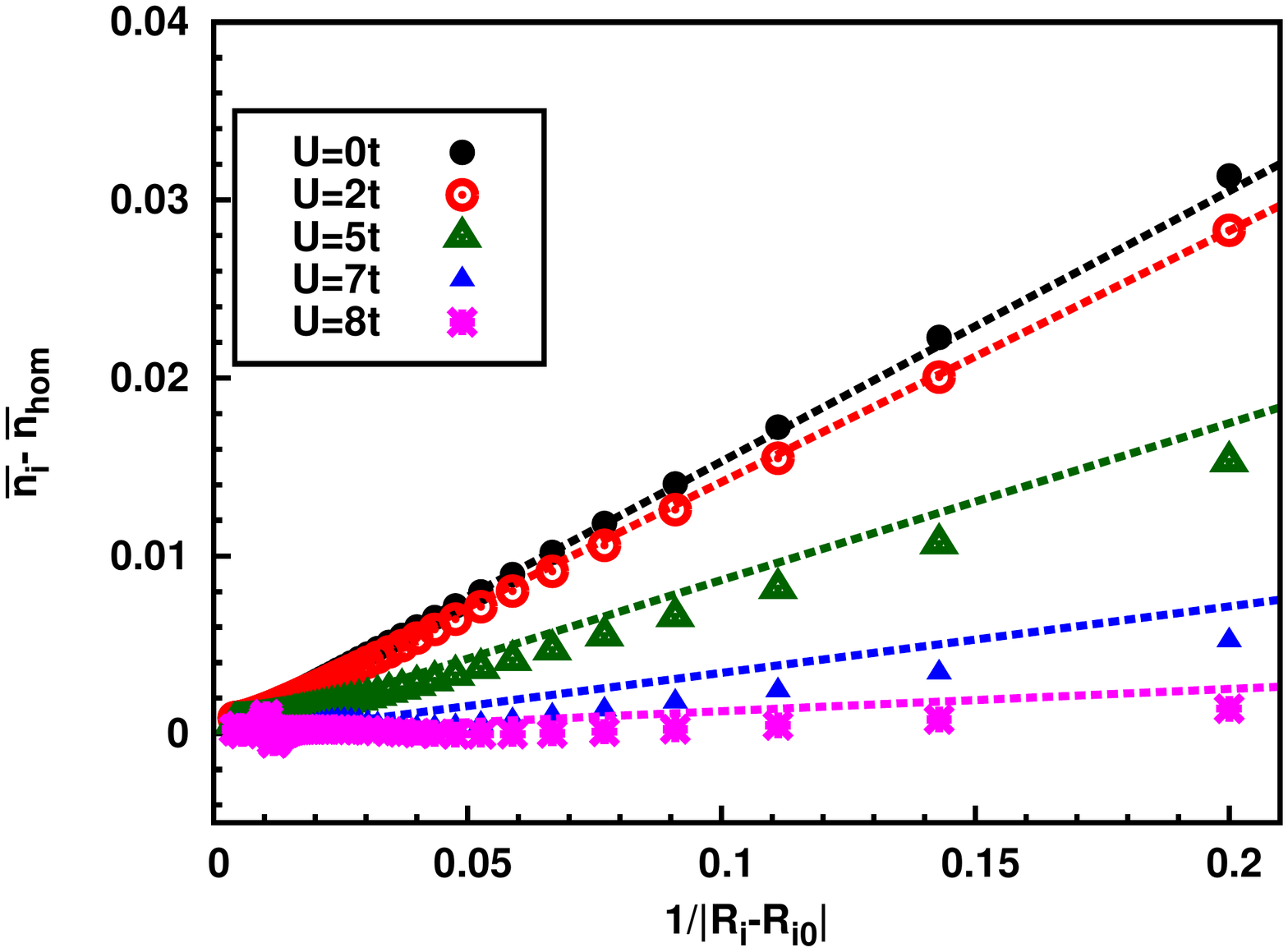} 		 
\caption{ Variation of the local density deviations as a function of inverse of the relative distance from the impurity site for the same system and method as in Fig.~\ref{fig8}.  }
\label{fig9}
\end{figure}

Similar to the analysis in the previous Section, in Fig.~\ref{fig9} we plot the local density deviations $|\bar{n}_{i}-\bar{n}_{hom}|$ as a function of the inverse of the relative distance from the impurity site, i.e. $1/|\vec{R}_{i}-\vec{R}_{i_{0}}|$.
The asymptotic linear decay is visible in agreement with Eq.~(\ref{friedel}) in the presence of interactions.

\begin{figure} [ht!]
\centering
\includegraphics[width=0.5\textwidth]{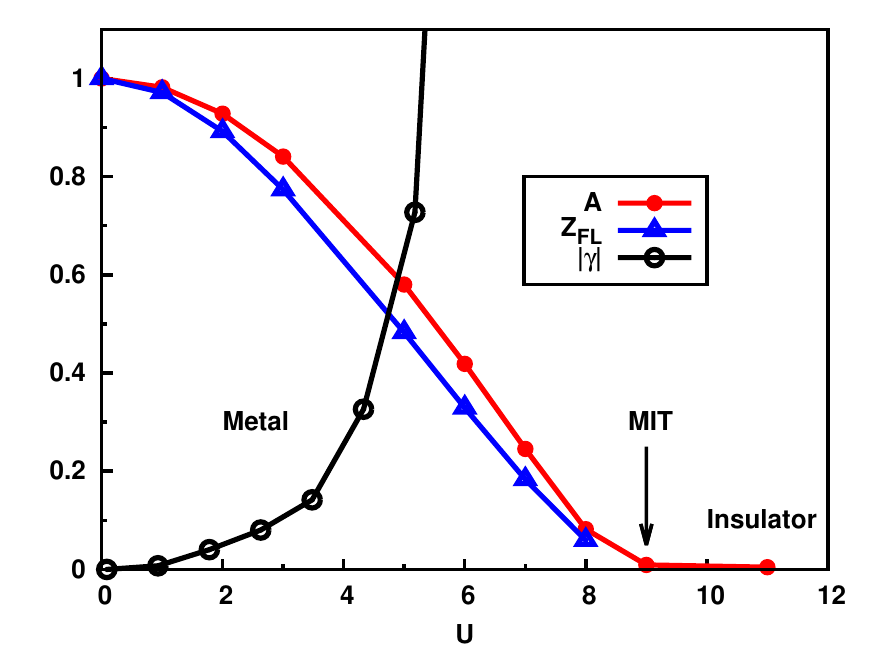} 		  
\caption{Variations  of the slope $A(U)/A(0)$  determined  from Fig.~\ref{fig9} (red curve), the Fermi liquid renormalization parameter (blue curve), and the inverse of quasiparticle life-time (black curve) as  functions of $U$.}
\label{fig10}
\end{figure}

The points shown in Fig.~\ref{fig9} follow the approximate linear rule $|\bar{n}_{i}-\bar{n}_{hom}|=Ax+B$, where $x=1/|\vec{R}_{i}-\vec{R}_{i_{0}}|$ and 
the parameters $A$ and $B$ are determined by fitting procedure.
In particular,  the slope parameter $A=A(U)$ describes changes in the FO amplitude with $U$.
We see that the slope $A(U)$ decreases with increasing $U$ and vanishes at the metal-insulator transition, cf. Fig.~\ref{fig10}, at $U_c\approx 9t$. 
The critical value $U_c$ determined from the $A(U)$ curve is in perfect agreement with the linearized DMFT \cite{Bulla00}. 
The value is larger than that obtained within CT-QMC because now the temperature is zero (strictly speaking $\beta=10000 t$ here).

Since a metallic system described with DMFT is within the Fermi liquid regime we expect that it must be fully quantified by the Fermi liquid renormalization parameter and by the life-time of quasiparticles. 
Both quantities are obtained by expanding the local self-energy at low frequencies, in the following way
\begin{equation}
\Sigma(\omega,U) \approx \alpha (U) \omega + i \gamma (U) \omega^2, 
\end{equation}
where parameters 
$\alpha (U)=(\partial {\rm Re} \Sigma(\omega,U)/\partial  \omega)|_{\omega=0}$ and $\gamma (U)=(\partial ^2 {\rm Im} \Sigma(\omega,U)/\partial \omega^2)|_{\omega=0}$ are determined numerically from the given NRG self-energy. 
In Fig.~\ref{fig10} we plot both the Fermi liquid renormalization factor $Z_{\rm FL}(U)=1/(1-\alpha(U))$ and the pre-factor $\gamma(U)$ in the inverse of the quasiparticle lifetime. 
The renormalization factor $Z_{\rm FL}(U)$ vanishes at the metal-insulator transition point $U_c$ whereas the coeffcient $\gamma(U)$ diverges there. 
We find that the FOs amplitudes $A(U)$ follow the behavior of $Z_{\rm FL}(U)$. 
This means that the renormalization of the quasiparticles wave functions is the primary source for damping of the FOs with increasing $U$. 
As expected $Z_{\rm FL}(U) =0$ at $U_c$ and FOs disappear.\\

As shown in \cite{Kittel} for the noninteracting systems, the particle density deviations away from the perturbing potential are given by $\Delta \bar{n}(r) = \lim _{R\rightarrow \infty} (R/\pi) \sum_{l=0}^{\infty} \int_0^{k_F} dk\int d \Omega (|\Psi_{kl}(r)|^2-|\Psi_{kl}^0(r)|^2)$, where $\Psi_{kl}(r)$ and $\Psi_{kl}^0(r)$ are partial components of the wave function in the spherical coordinates in the presence and the absence of the impurity, respectively. Taking into account the interaction effects in the Fermi liquid picture, the wave functions are renormalized (multiplied) by the square root of the renormalization factor $\sqrt{Z_{\rm FL}(U)}$. Thus we expect that $\Delta \bar{n}(r) \sim Z_{\rm FL}(U)$ in good agreement with our numerical findings.

The screening charge $Z$, defined in Eq.~\ref{screen1}, is shown in Fig.~\ref{fig11}. 
The repulsive  $V_0>0$ potential leads to lowering the number of particles in the system whereas the attractive $V_0<0$  potential yields this number to increase. 
There is a perfect mirror symmetry between these two regimes as is seen in the upper panel in Fig.~({\ref{fig11}) for different $U$. 
With increasing the interaction the screening of the impurity is weaker, i.e. the number of removed charge is smaller, as is presented in the upper and lower panels of Fig.~\ref{fig11}.

Although the screening charge $|Z|$ is a decreasing function of $U$ it remains finite in the Mott insulator where FOs are absent, cf. the lower panel of Fig.~\ref{fig11}. 
The reason is that at the impurity site $\vec{R}_{i_0}$ and in its vicinity the particle density is different from $\bar{n}_{\rm hom}$ for any finite $V_0$. 
Therefore, the screening charge $Z$, which counts the particle deviations for all lattice sites in Eq.~\ref{screen1}, is finite even if FOs in the asymptotic regime (very far from the impurity site) are absent.

\begin{figure} [ht!]
\centering
\includegraphics[width=0.5\textwidth]{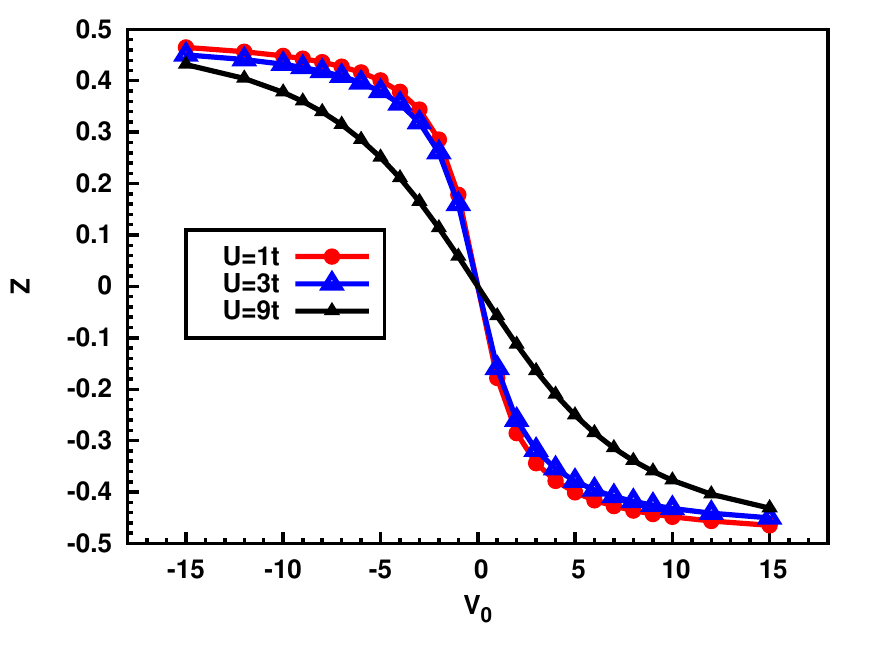} 		 
\qquad
\includegraphics[width=0.5\textwidth]{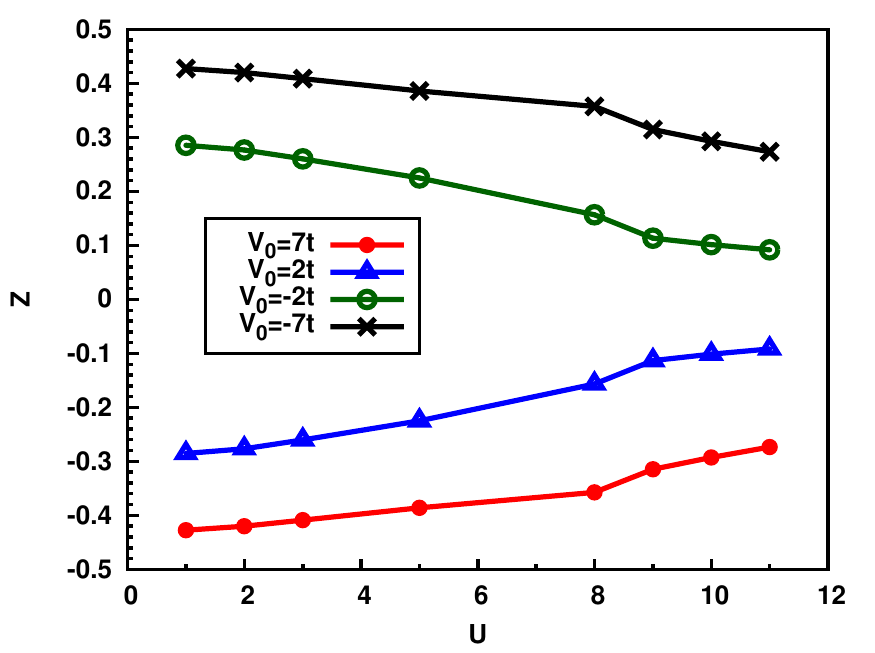} 
\caption{ Screening charge $Z$ determined from Eq.~(\ref{screen1}) as a function of the impurity potential $V_0$ for selected $U$ (upper panel) and as a function of the interaction $U$ for $V_0=\pm 2t$ and $\pm7t$ (lower panel). The other system parameters are the same as in Fig.~\ref{fig8}. }
\label{fig11}
\end{figure}

 \subsection{Two dimensional square lattices}
 
 The FOs in the two dimensional lattice with the HSEA are presented in Figs.~\ref{fig12}~and~\ref{fig13}. 
 The lattice size is $31\times31$  and the impurity potential $V_0=24t$ is located in the center. 
 The main panels of Figs.~\ref{fig12}~and~\ref{fig13} show two-dimensional color maps of FOs seen in the particle density. 
 In the insets the FOs are shown along the vertical line crossing the impurity site. 
 The FOs are not spherically symmetric as in a free space but possess the square lattice symmetry. 
 Due to constructive interference of oscillatory waves along horizontal and vertical directions the strongest FOs are observed along these lines. 
 The amplitudes of FOs are much smaller as compared to those in one-dimensional cases because of the damping factor $1/r^2$. 
 It is also clearly seen that by increasing the interaction the amplitudes of the oscillations are weaker and as the system becomes an insulator the FOs far from the impurity are absent. 
In insulator there are still  visible deviations in the density of particles in the neighborhood of the impurity site.
In order to highlight the behavior of FO in vicinity of the impurity potential the value of the density at the impurity site has been substituted by the second smallest value calulated in the system. Without this correction in the plots the impurity contribution would overshadow any oscillations in $n_{i}$ due to their power law behavior with distance as shown in Fig.~\ref{fig14}.

\begin{figure*}
    \centering
    \begin{minipage}{0.45\textwidth}
        \centering
        \includegraphics[width=0.9\textwidth]{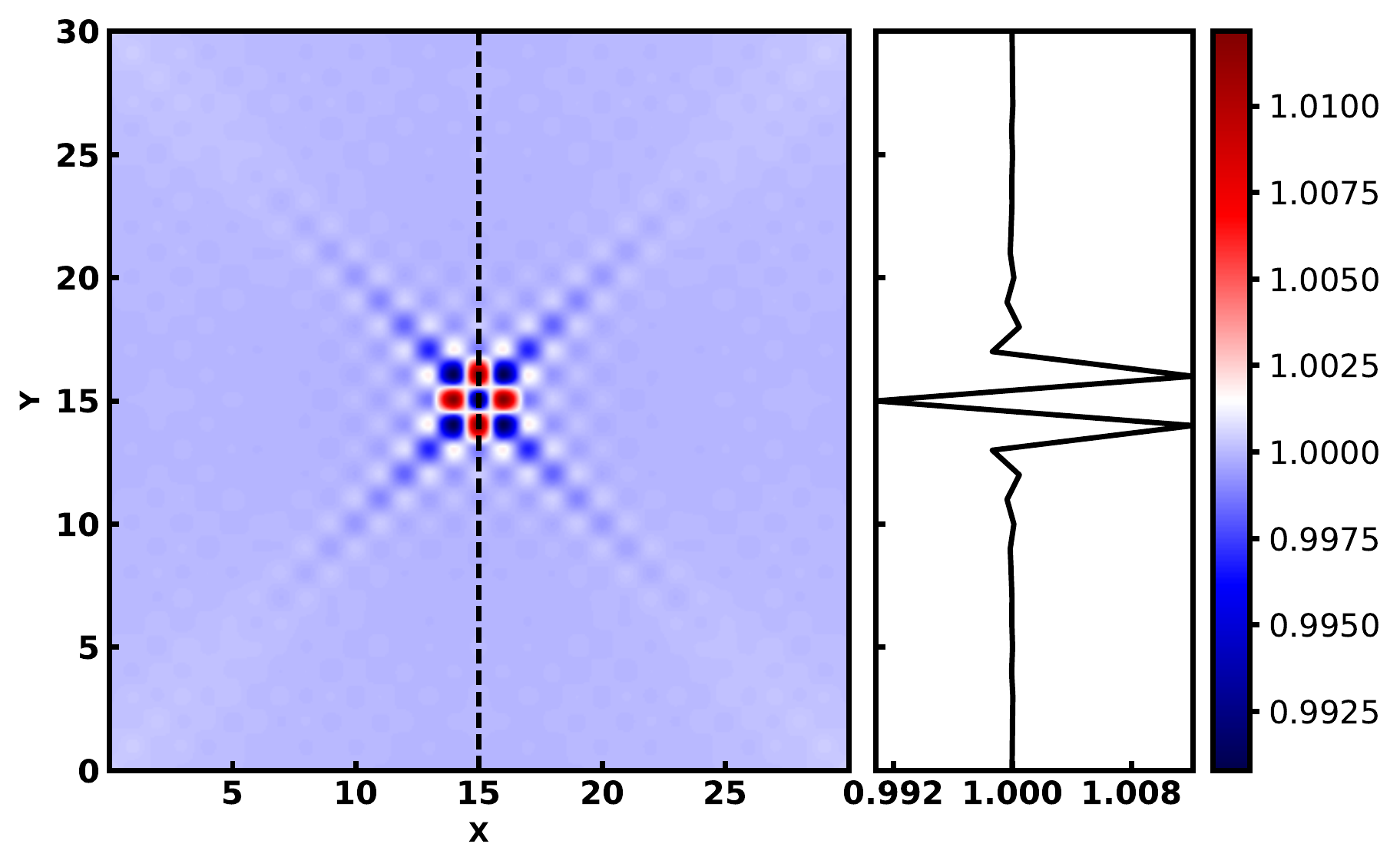} 
    \end{minipage}\hfill
    \begin{minipage}{0.45\textwidth}
        \centering
        \includegraphics[width=0.9\textwidth]{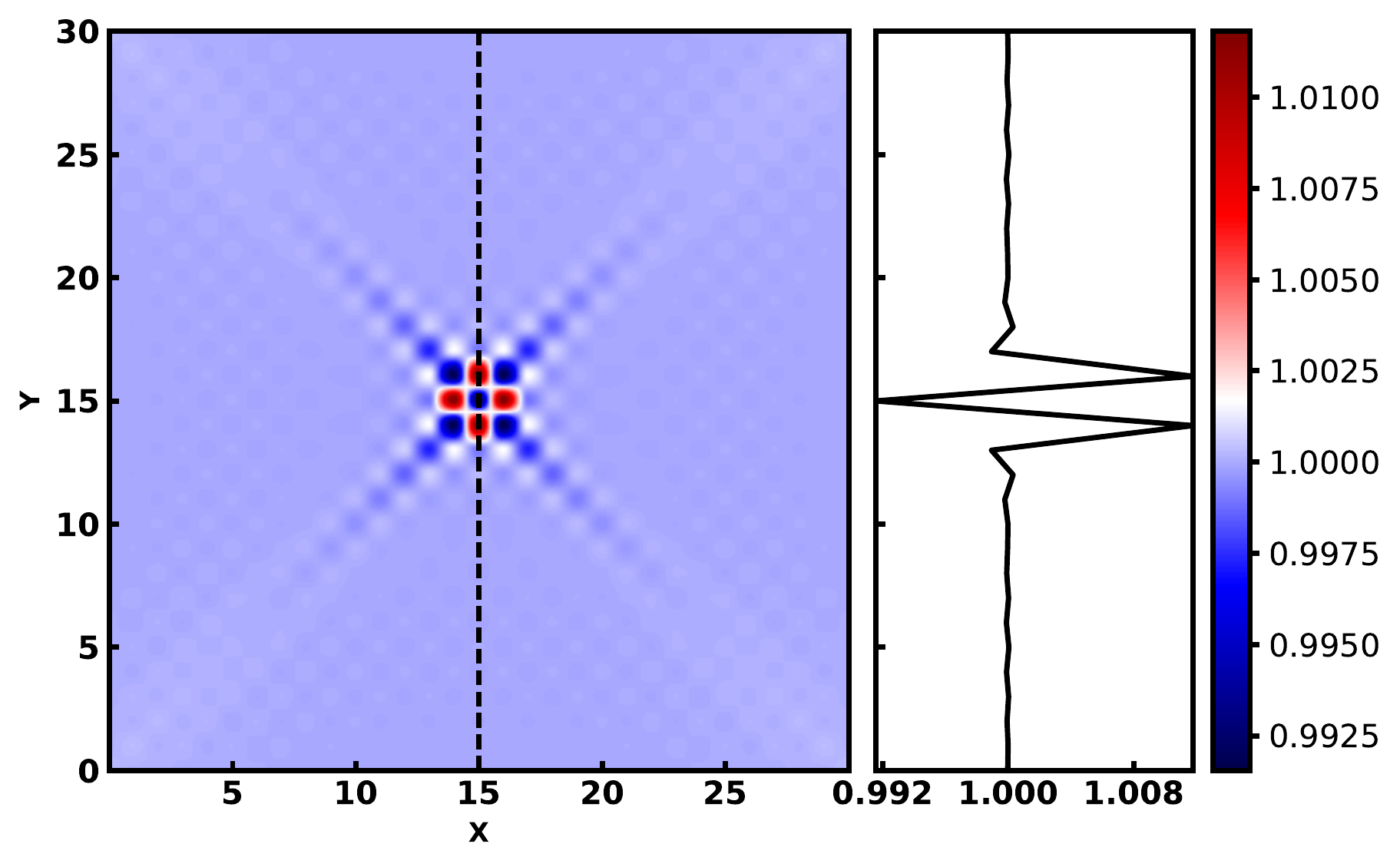} 
    \end{minipage}
\caption{Friedel oscillations in particle densities of the Hubbard model in two-dimensional square lattice with $31\times 31$ sites. The impurity potential $V_0=24t$ is located in the center at $\vec{R}_0=(15a,15a)$ and the interaction $U=0$ (left panel) and $2t$ (right panel).  The insets show FOs along vertical line crossing the impurity site. The color scale is spanned in between the highest and lowest values of density in the system. The color scale changes for different U's since the minimal value of the density increases with U as shown in the insets.}
\label{fig12}
\end{figure*}

\begin{figure*}
    \centering
    \begin{minipage}{0.45\textwidth}
        \centering
        \includegraphics[width=0.9\textwidth]{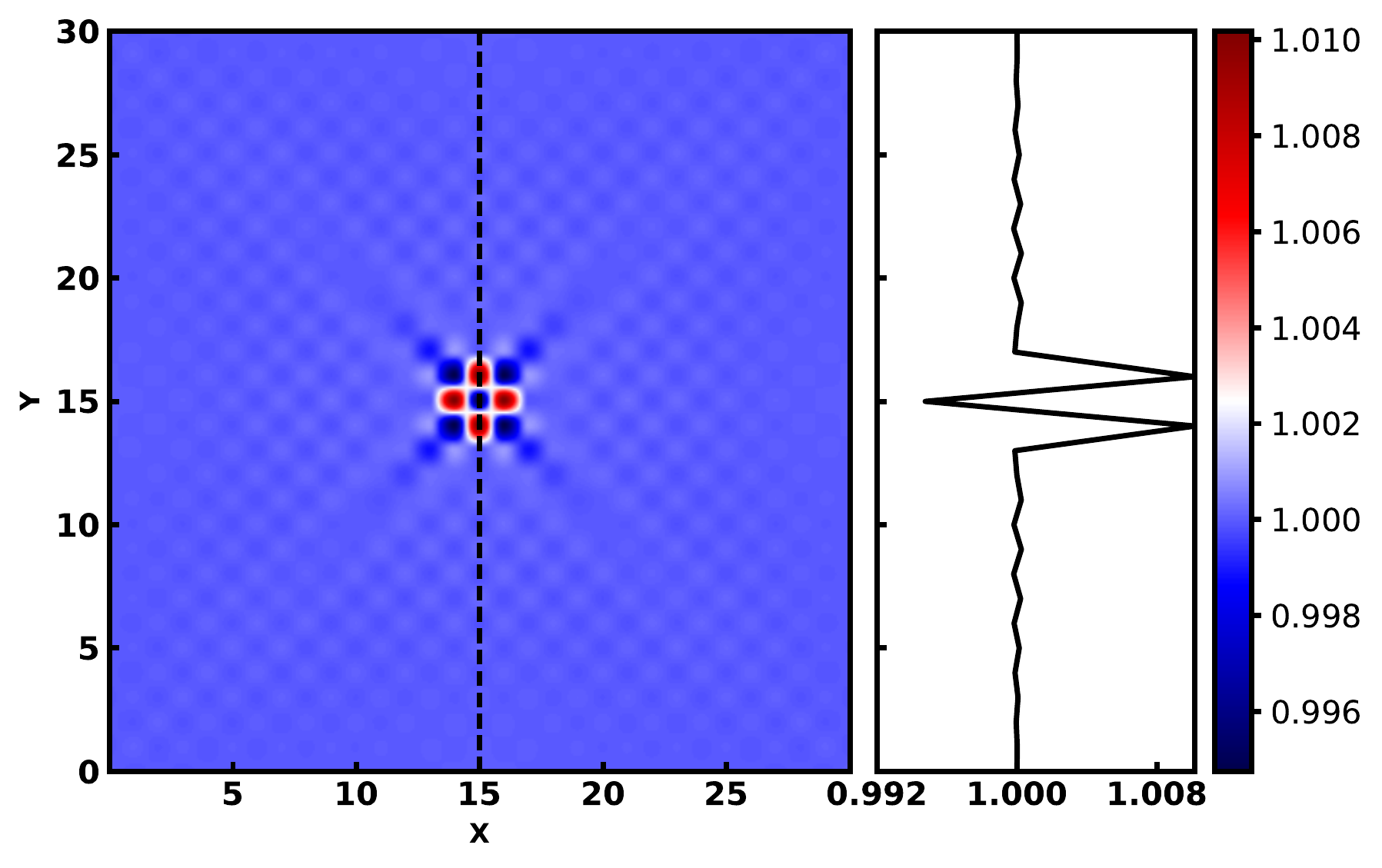} 
    \end{minipage}\hfill
    \begin{minipage}{0.45\textwidth}
        \centering
        \includegraphics[width=0.9\textwidth]{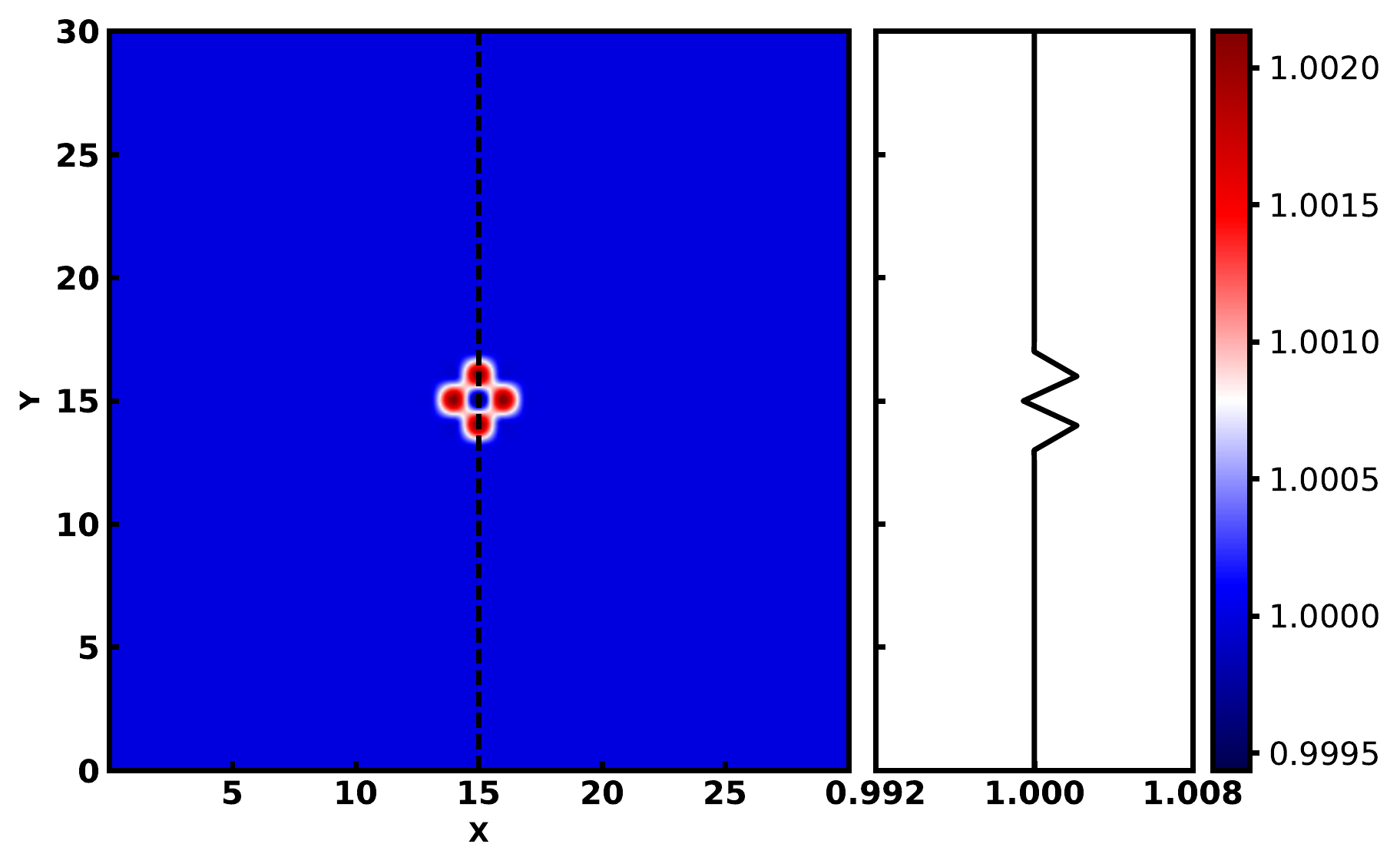} 
    \end{minipage}
\caption{Friedel oscillations in particle densities of the Hubbard model in two-dimensional square lattice with $31\times 31$ sites. The impurity potential $V_0=24t$ is located in the center at $\vec{R}_0=(15a,15a)$ and the interaction $U=5t$ (left panel) and $12t$ (right panel).  The insets show FOs along vertical line crossing the impurity site and the color bars have the same legend as in Fig.~\ref{fig12}. }
\label{fig13}
\end{figure*}

\begin{figure} 
\centering
\includegraphics[width=0.5\textwidth]{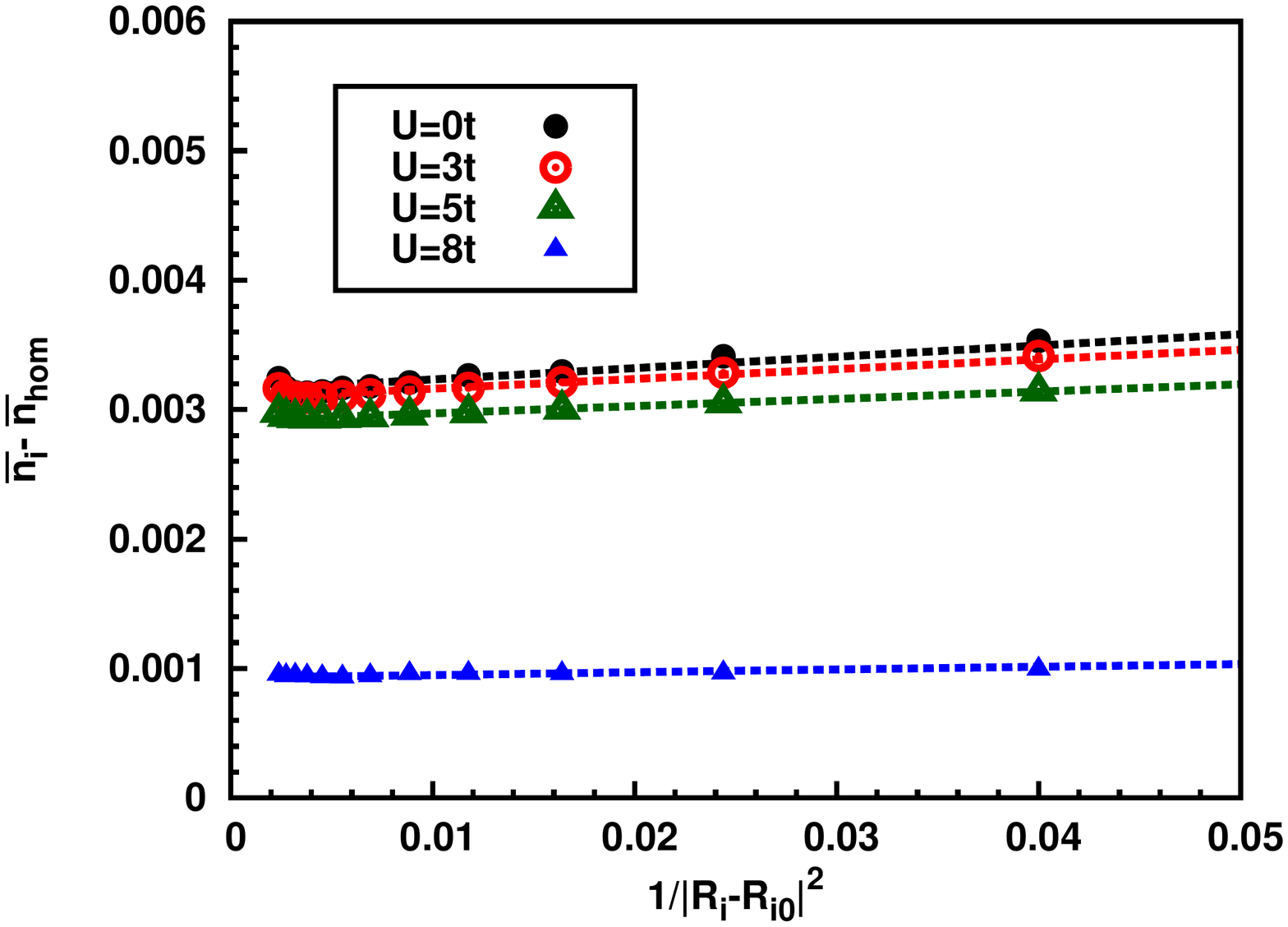} 		 
\caption{ Variation of the density deviations as a function of the square of inverse of the relative distance from the impurity site for the same system and method as in Fig.~\ref{fig12}~and~\ref{fig13}.  }
\label{fig14}
\end{figure}

The densities shown in Figs.~\ref{fig12}~and~\ref{fig13} follow the approximate quadratic rule $|\bar{n}_{i}-\bar{n}_{hom}|=Ax^2+B$, where $x=1/|\vec{R}_{i}-\vec{R}_{i_{0}}|$ and 
the parameters $A$ and $B$ are determined by fitting procedure, cf. Fig.~\ref{fig14}. 
Here we show results along a line parallel to the diagonal one and shifted by one lattice constant. 
We see that the slope $A(U)$ decreases with increasing $U$ and vanishes at the metal-insulator transition, cf. Fig.~\ref{fig15}, at $U_c\approx 12t$, again, in perfect agreement with the linearized DMFT \cite{Bulla00}.

In Fig.~\ref{fig15} we plot both the Fermi liquid renormalization factor $Z_{\rm FL}(U) $ and the pre-factor $\gamma (U)$ in the inverse of the quasiparticle lifetime for the  Hubbard model in two-dimensions. 
The renormalization factor $Z_{\rm FL}(U) $ vanishes at the metal-insulator transition point $U_c$ but  $\gamma (U) $ diverges there. 
Again we see that the FOs amplitudes $A(U)$ follow the behavior of $Z_{\rm FL}(U) $ and the renormalization of the quasiparticles wave functions is the primary factor for damping of the FOs with increasing $U$.

 \begin{figure} [h!]
\centering
\includegraphics[width=0.5\textwidth]{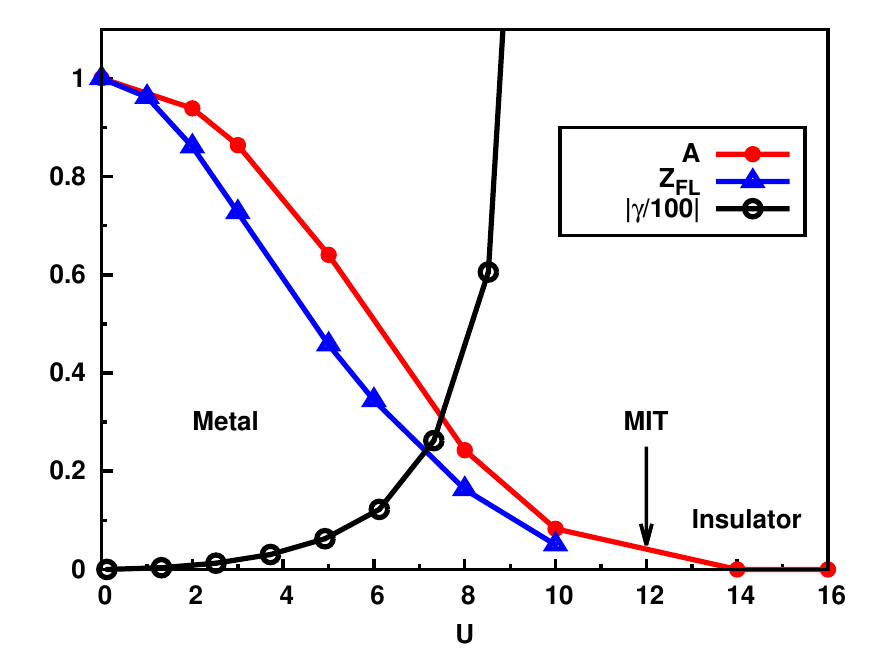} 		  
\caption{Variations  of the slope $A(U)/A(0)$  determined  from Fig.~\ref{fig13} (red curve), the Fermi liquid renormalization parameter (blue curve), and the inverse of quasiparticle life-time (black curve) as  functions of $U$.}
\label{fig15}
\end{figure}

 The screening charge $Z$ is shown in Fig.~\ref{fig16}. 
The repulsive  $V_0>0$ potential $Z$ is negative because particles are repelled from the system and for $V_0<0$ the situation is perfectly reflected. 
With increasing the interaction the screening decreases, i.e. the number of removed particles is smaller, as is shown in both  panels of Fig.~\ref{fig16}. 
The screening charge $|Z|$ remains finite in the Mott insulator where FOs are absent because  the particle density is different from $\bar{n}_{\rm hom}$ for any finite $V_0$ very close to the impurity site, see Fig.~\ref{fig13}. 
 
\begin{figure} [h!]
\centering
\includegraphics[width=0.5\textwidth]{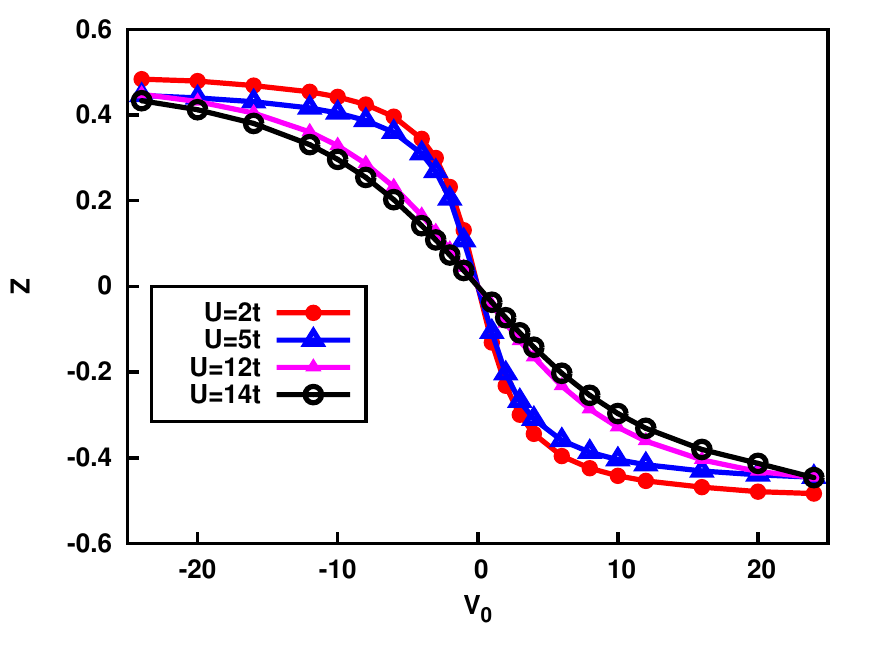} 		 
\qquad
\includegraphics[width=0.5\textwidth]{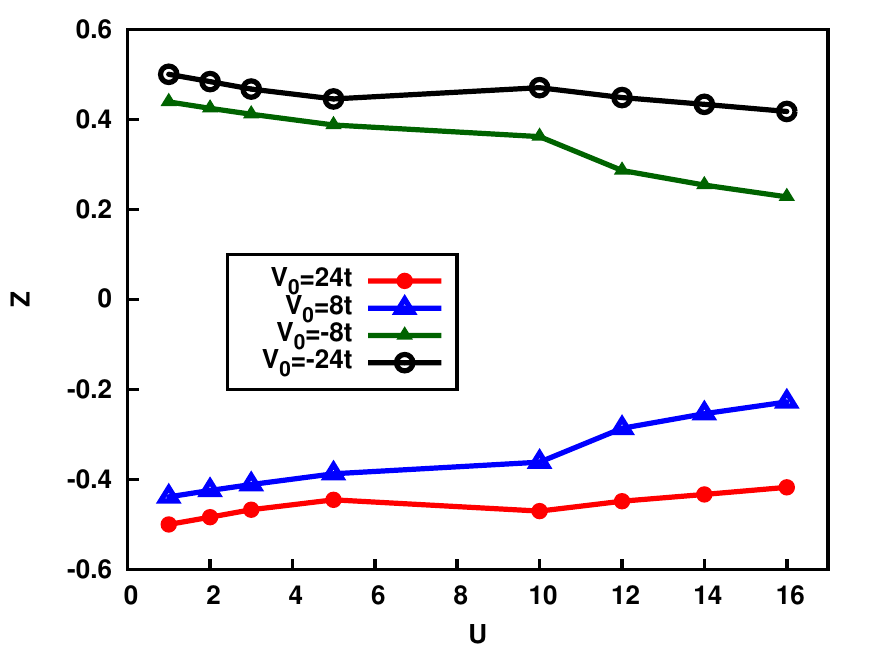} 
\caption{ Screening charge $Z$ determined from Eq.~(\ref{screen1}) as a function of the impurity potential $V_0$ for selected $U$ (upper panel) and as a function of the interaction $U$ for $V_0=\pm 8t$ and $\pm 24t$ (lower panel). The other system parameters are the same as in Fig.~\ref{fig12}. }
\label{fig16}
\end{figure}
 
 Finally, we compare the  screening charge $Z$ with the Fermi liquid renormalization factor $Z_{\rm FL}$ for the same interaction $U$ in Fig.~\ref{fig17}. In this case $Z$ is calculated for $V_{0}= 7t (8t)$ for 1d (2d)
 systems. We find very good linear dependence of these quantities both in one and two dimensional lattices for our chosen $V_{0}$.

\begin{figure} [h!]
\centering
\includegraphics[width=0.5\textwidth]{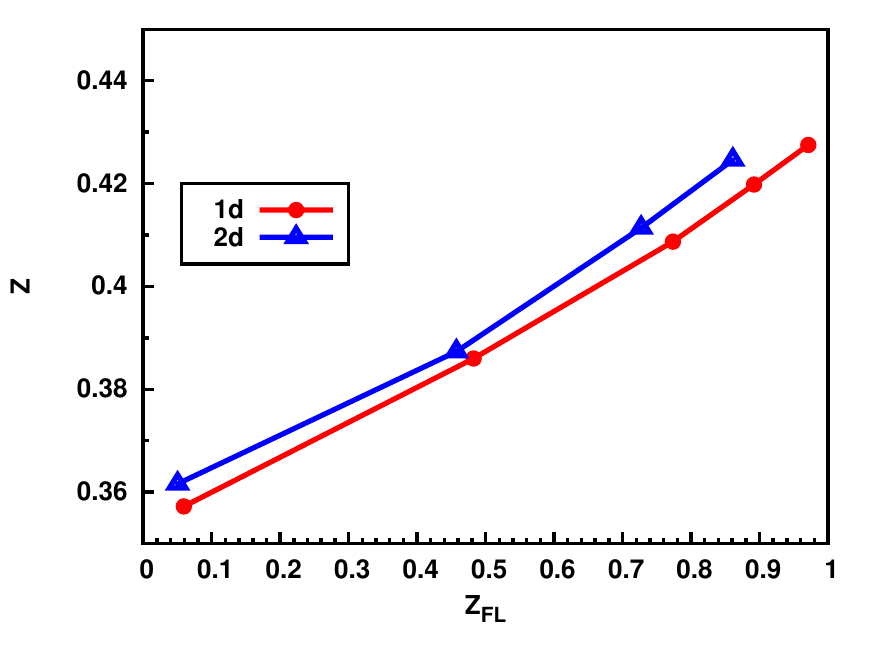} 		  
\caption{Comparision of the screening charge $Z$ with the Fermi liquid renormalization factor $Z_{\rm FL}$ in one and two dimensional systems with $V_0=7t$ and $8t$ respectively. }
\label{fig17}
\end{figure}

 \section{Conclusions and  outlook}
\label{summary}

We investigated the Friedel oscillations and screening effects around an impurity potential within the Hubbard model in one and two dimensional lattices. 
We solved the Hubbard model within the real-space dynamical mean-field theory at finite and zero temperatures by using continuous time Monte-Carlo simulations and homogeneous self-energy approximations with numerical renormalization group, respectively. 

In one and two-dimensional metallic, Fermi liquid regime the Friedel oscillations are damped by increasing the interaction but their decaying pattern $1/r^d$, with $d=1$ and $2$, is the same as in the non-interacting ideal Fermi gas. 
We found that decaying of amplitude oscillations follows the behavior of the Fermi liquid renormalization factor, which decreases with increasing the interaction. 
The same holds for the screening charge. 
We also observed that the life-time of Fermi liquid quasiparticles, which vanishes at the metal-insulator transition point, does not play any essential  role in understanding the behavior of Friedel oscillations. 
In the Mott insulating phase the Friedel oscillations, very far from the impurity potential, are absent. 
Only very close to the impurity site there are deviations in the density with respect to homogeneous systems. 
The screening charge remains finite in the insulator. 
We found very good agreement on a qualitative and even quantitative level between the exact Monte-Carlo simulations and the homogeneous self-energy approximations. 
We conclude that the homogeneous self-energy approximations, which is much cheaper in a computational time cost,  is a reliable approximation for the present problems. 

Interesting behavior and Friedel like oscillations are  seen in the local spectral functions, which we determined as well. 
We also obtained initial results for Friedel oscillations in cases of few impurities in the system. 
Interesting interference fringes are found there. 
These new results  deserve a separate publication. 

The present study is planned to be extended into three dimensional systems and systems with larger lattice sites to clearly see the asymptotic $1/r^3$ behavior in the amplitude decaying of Friedel oscillations. 
For this the real-space mean-field theory either within the Monte-Carlo simulations or within the numerical renormalization group method should be parallelized and computational task should be split on many computing platforms.

\begin{acknowledgments}
We thank for fruitful  discussions with D.~Vollhardt and J.~Panas.
In years 2011-15 this work was supported by Foundation for Polish Science (FNP) through the TEAM/2010-6/2 project, co-financed by the EU European Regional Development Fund.
Support by the Deutsche Forschungsgemeinschaft through TRR 80 is also gratefully acknowledged. BC acknowledges discussions with J. Kolorenc and financial support from the Czech Academy of Sciences in the year 2018.\\
\end{acknowledgments}


\begin{thebibliography}{}
\bibitem {1D}
According to the exact Bethe ansatz  solution of this model in one-dimension and at half-filling there is a Mott insulating state for any finite interaction \cite{Lieb68}. 
Moreover, away from half-filling  the Fermi liquid picture breaks there and the electrons are in a Luttinger liquid state. 
DMFT provides an  exact solution of the Hubbard model in the infinite dimension \cite{Metzner89}. 
However, at finite dimensions DMFT approximates the self-energy by  the momentum independent one and, therefore,  reproduces features of Fermi liquid even in one dimension. 
In the paper we study the one-dimensional Hubbard model  within R-DMFT only to gain an initial insight about and make a visualization of FOs in interacting systems.
All results for one- and two-dimensional lattices are consistent with the Fermi liquid picture.
\bibitem{note-phase-shift} We have checked that away from the half-filling this invariance does not hold any more.
\end{thebibliography}
\end{document}